**Explaining Ground-Motion Residuals at Two Strong-Motion Stations in Southeastern New York: Sediment Resonance and Topographic Amplification**

Aser Abbas*, Patrick Daniele, James Kaklamanos, Laurie Baise, Kyle Cannon, and Ellie Meyer

## ABSTRACT

Sites hosting strong-motion stations in the Central and Eastern United States are commonly characterized using proxy-based parameters, limiting the ability to identify the physical causes of large ground-motion residuals. This study investigates the amplification mechanisms at two New York strong-motion stations that recorded some of the largest positive 1-Hz pseudospectral-acceleration residuals during the 2024 $M_w$ 4.8 Tewksbury, New Jersey, earthquake. Station N4.N62A is located at Caumsett State Historic Park in the Atlantic Coastal Plain on Long Island, whereas station LD.PAL is located at the Lamont-Doherty Earth Observatory atop the Palisades cliffs along the Hudson River. Detailed geophysical site characterization at both sites combined active-source and ambient-noise surface-wave testing with horizontal-to-vertical spectral ratio (HVSR) measurements. At Lamont-Doherty, additional ambient-noise arrays were deployed across the Hudson River Palisades ridge to evaluate topographic amplification. At Caumsett, surface-wave testing resolved a thick sedimentary column overlying a major impedance contrast, with a spatially representative fundamental site frequency of 0.93 Hz. Amplification in the same frequency range was also independently evident in observed amplifications computed from site-to-site (S2S) residuals. The fundamental-resonance band also encompasses the 1-Hz frequency at which the large Tewksbury residual was observed, supporting deep sediment resonance as the dominant mechanism. At Lamont-Doherty, surface-wave testing indicated hard-rock conditions with a thin, laterally variable sediment cover. The ambient-noise arrays showed repeatable, directional amplification near 2 Hz at LD.PAL, and the earthquake-based S2S amplifications show a peak between approximately 0.6 and 2 Hz, a pattern reproduced by neither the transfer functions nor the ergodic linear amplification models. These observations support topographic amplification as the primary mechanism at LD.PAL. The results demonstrate that comparable positive residuals at stations 38 km apart can arise from fundamentally different site-effect mechanisms and highlight the value of targeted field characterization for informing more physically based, non-ergodic ground-motion models.

*Assistant Professor, Dept. of Civil and Environmental Engineering, University of Rhode Island, 2 East Alumni Avenue, FCAE Room 209, Kingston, RI 02881, USA, aser.abbas@uri.edu

## 1. INTRODUCTION

Persistent residuals between strong-motion recordings and ground-motion model (GMM) predictions remain a central challenge in seismic hazard analysis (Stewart et al., 2020; Meyer et al., 2026). These residuals are often associated with site effects: the modification of seismic waves by geologic materials and topography encountered as the waves propagate toward the ground surface (Borcherdt, 1970; Aki, 1993). Improving the physical basis of GMMs therefore requires moving away from ergodic site terms, which represent an individual site by the average response of many broadly similar sites, toward non-ergodic models that represent repeatable site-specific effects explicitly (Kramer and Stewart, 2025). In the Central and Eastern United States (CEUS), however, the site characterization needed to support this transition is rare; only about 6% of stations in the NGA-East database (Goulet et al., 2021a) have a time-averaged shear-wave velocity in the upper 30 m ($V_{S30}$) derived from direct field measurements, with the remainder relying on slope- or geology-based proxies. Without measurements at the stations themselves, elevated residuals cannot be confidently traced to the physical mechanisms that produce them.

Local ground response and topographic amplification are two distinct site-effect mechanisms (Kramer and Stewart, 2025). Both are relevant in the northeastern United States, where glacial deposits, sediment-filled valleys, coastal deposits, and pronounced bedrock topography are widespread (Pontrelli et al., 2023). This study investigates these two mechanisms at two strong-motion stations in southeastern New York.

Local ground response occurs when low-velocity sediments modify ground motions through impedance and resonance effects (Idriss and Seed, 1968; Borcherdt, 1970). This mechanism is particularly consequential in the northeastern United States. Glacial and coastal sediments overlying competent bedrock produce strong impedance contrasts, and the resulting resonant amplification at the fundamental site frequency can reach a factor of 10, which exceeds the values currently considered in seismic design codes (Baise et al., 2016). In these environments, amplification is controlled by sediment thickness and impedance contrasts rather than by the near-surface velocities captured in $V_{S30}$ alone. Large impedance contrasts are especially common in CEUS site profiles, and the dependence of site response on $V_{S30}$ there is correspondingly weaker than in the western United States (Parker et al., 2019). Reflecting these limitations, the 2023 National Seismic Hazard Model incorporated sediment-thickness-dependent adjustments for the Atlantic and Gulf Coastal Plains, while alternative site parameters are being considered to improve future versions of these models (Petersen et al., 2024). Progress toward such site terms, however, is limited by the scarcity of measured deep velocity profiles in the CEUS (Gann-Phillips et al., 2024, 2026). New measurements at stations with observed high site-to-site (S2S) residuals provide better-constrained benchmark datasets for evaluating site amplification models and for developing alternative site terms for GMMs.

Topographic effects occur where irregularities in the ground surface geometry, such as ridges and slopes, redistribute seismic energy through wave scattering, diffraction, and focusing (Ashford et al., 1997). This site effect mechanism is similarly consequential but even less well represented in current practice. Earthquake observations have repeatedly shown that ridge crests and hilltops experience amplified shaking, contributing to disproportionate damage on topographic highs (Massa et al., 2010). Experimental studies have documented crest amplifications of roughly 4 to 4.5, with the strongest motions polarized perpendicular to ridge elongation (Massa et al., 2010; Pischiutta et al., 2010). Nevertheless, the GMMs underlying the 2023 National Seismic Hazard Model do not explicitly incorporate topographic effects, so shaking amplified by surface geometry is absorbed into unexplained residual variability (Petersen et al., 2024; Meyer et al., 2026). Terrain-based proxies such as curvature and relative elevation offer a promising path toward filling this gap, but they require validation against well-constrained field observations at real ridge sites (Maufroy et al., 2015; Rai et al., 2017). Field measurements at topographically complex stations are therefore needed on two timescales. In the long term, they can support the development of GMMs that represent topographic amplification explicitly. In the short term, they can identify stations whose recordings are influenced by topography and should not be used blindly in GMM development (Rischette et al., 2026).

The two stations examined here are N4.N62A at Caumsett State Historic Park (Atlantic Coastal Plain, Long Island) and LD.PAL at the Lamont-Doherty Earth Observatory (atop the Palisades cliffs along the Hudson River). At each site, geophysical site characterization is combined with observed and predicted amplifications to identify the dominant mechanism. The remainder of this article is organized as follows. Section 2 describes observations from the 2024 Tewksbury, New Jersey, earthquake, which had a moment magnitude (M*w*) of 4.8, and explains the rationale for selecting these two stations. Section 3 presents the site-characterization methods and results for Caumsett State Historic Park and evaluates deep sediment resonance as the primary amplification mechanism. Section 4 presents the investigation at the Lamont-Doherty Earth Observatory and evaluates topographic amplification as the primary mechanism. Section 5 summarizes the principal findings and highlights their implications for site characterization and ground-motion modeling in the CEUS.

## 2. THE 2024 TEWKSBURY EARTHQUAKE AND STUDY SITES

On April 5, 2024, an $M_w$ 4.8 earthquake occurred near Tewksbury, New Jersey, approximately 65 km west of New York City (USGS, 2024). Although moderate in magnitude, the event was the largest instrumentally recorded earthquake in New Jersey since 1900 (Han et al., 2024), and the low attenuation characteristic of the CEUS allowed its ground motions to be recorded across the northeastern United States. We evaluated residuals between recorded 1-Hz pseudospectral accelerations and predictions from the weighted median of the NGA-East GMMs (Goulet et al., 2021b) and identified broad spatial variability across the region. Similar results were found by Meyer et al. (2025) using the Atkinson and Boore (2006) GMM. In general, stations east of the Hudson River recorded stronger motions than predicted, whereas stations farther west recorded weaker motions than predicted, as illustrated in Figure 1b. In prior work using ground motions from across the CEUS, Meyer et al. (2026) showed that regional variability in CEUS linear site response can be represented by a two-parameter linear site term based on adjusted physiographic province and sediment thickness. As shown in Figure 1a, this term results in significant deamplification in the New England adjusted province and significant amplification in the Embayed province (the section of the Atlantic Coastal Plain that includes Long Island). These regional trends represent the repeatable component of station response, but individual stations can still exhibit anomalously large residuals during a single event.

Two stations that recorded among the largest positive 1-Hz residuals in the region were selected for field investigation (Figure 1b). The first is station N4.N62A at Caumsett State Historic Park on Long Island, New York, within the Embayed adjusted physiographic province. The second is station LD.PAL at the Lamont-Doherty Earth Observatory along the Hudson River Palisades, near the boundary between the Piedmont and New England adjusted physiographic provinces. Two further considerations motivated this selection. First, the two sites occupy sharply contrasting geologic settings that suggest different candidate amplification mechanisms: Caumsett is underlain by thick unconsolidated coastal-plain sediments, whereas LD.PAL has minimal sediment cover and sits on the Palisades ridge, approximately 170 m inland from its eastern crest. Second, the stations are located approximately 38 km apart and lie within the greater New York City metropolitan region, where the potential consequences of amplified shaking are especially significant. As described later in this paper, we paired the site characterization data with an evaluation of linear site response using several independent methods to better understand the reasons for the positive residuals.

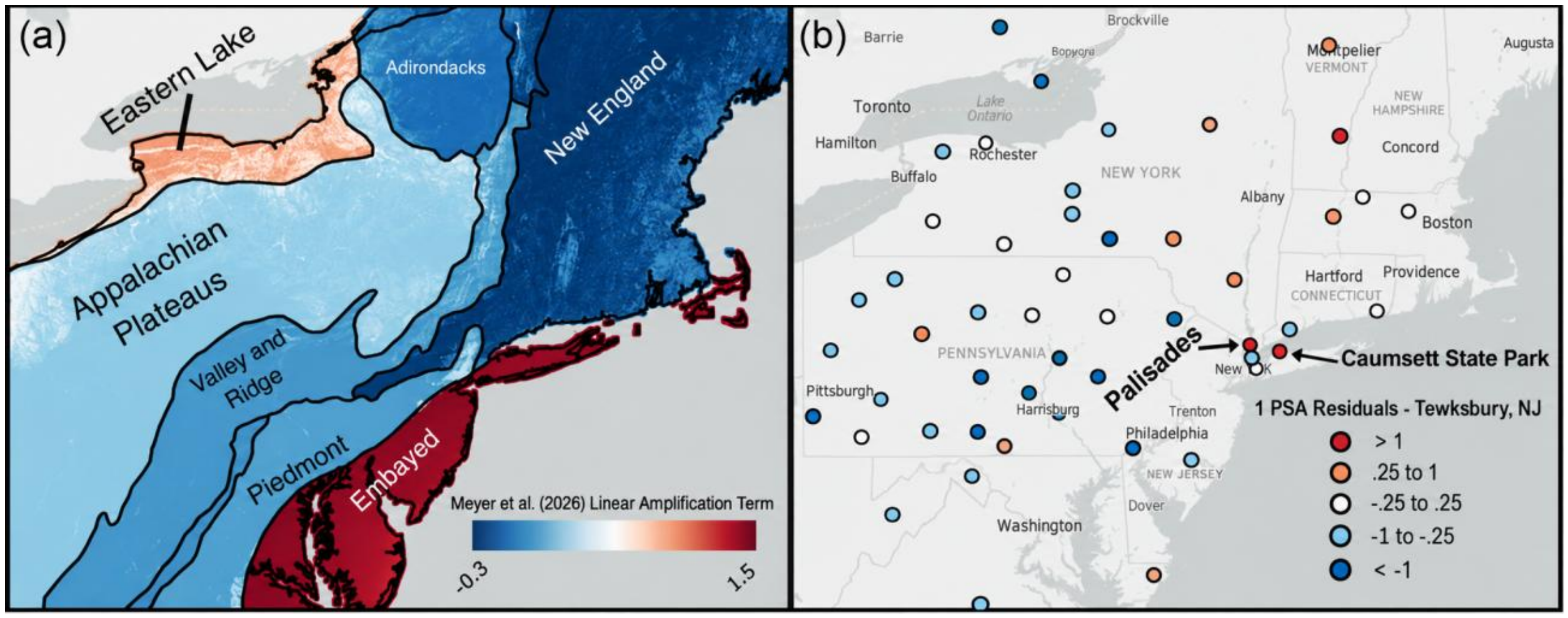

*Figure 1. (a) mapped Meyer et al. (2026) linear amplification term at 1-Hz PSA based on adjusted physiographic province and sediment thickness using NGA-East Median GMM; (b) Regional distribution of 1-Hz pseudospectral acceleration residuals for the 2024 $M_w$ 4.8 Tewksbury, New Jersey, earthquake, relative to the weighted median of the NGA-East GMMs (Goulet et al., 2021b). Arrows identify Caumsett State Historic Park (station N4.N62A) and the Lamont-Doherty Earth Observatory (station LD.PAL, labeled as Palisades in the figure). The red symbols indicate residuals greater than 1, placing both stations among those with the largest positive residuals in the region.*

## 3. CAUMSETT STATE HISTORIC PARK

### 3.1 Site Description and Surface-Wave Testing

As described in Section 2, station N4.N62A is located within Caumsett State Historic Park on the northern shore of Long Island, where unconsolidated glacial and coastal-plain sediments overlie crystalline bedrock (Smolensky et al., 1989). The site was characterized using surface-wave testing, which proceeds in three stages: data acquisition, processing, and inversion. These stages are described in turn below.

***Data Acquisition***

Data were acquired during two field campaigns, the first of which was conducted on July 28 and 29, 2025, and the second on November 29, 2025. Testing consisted of active-source multichannel analysis of surface waves (MASW; Park et al., 1999), passive microtremor array measurements (MAM; Tokimatsu, 1997; Okada, 2003; Ohrnberger et al., 2004), and horizontal-to-vertical spectral ratio (HVSR) measurements (Figure 2). MASW testing was performed using two linear arrays of 24 geophones each (Geospace Technologies GS-ONE), spaced 2 m apart, for a total array length of 46 m. Vertical 4.5-Hz geophones were used to record Rayleigh waves, while horizontal 4.5-Hz geophones, oriented transverse to the array, were used to record Love waves. Rayleigh-wave energy was generated by striking a round aluminum strike plate vertically with a 7.3-kg sledgehammer, while Love-wave energy was generated by striking a weighted shear-traction plank oriented perpendicular to the array with the same sledgehammer. Both source types were applied at eight shot locations, offset 5, 10, 15, and 20 m beyond each end of the array, to provide forward and reverse shots. Five hammer blows were stacked at each location to increase the signal-to-noise ratio (Foti et al., 2018). Each record was 2.0 s long, with a 0.5-s pre-trigger delay and a 1-ms sampling rate.

MAM testing was conducted using three-component broadband seismometers with a flat frequency response from 0.2 to 150 Hz (SmartSolo IGU-BD3C-5). The seismometers were oriented toward magnetic north, buried, and covered with weighted buckets to provide adequate ground coupling and reduce wind-induced noise. Ambient vibrations were recorded using three circular arrays (C50, C300, and C2150) and one concentric triangular array (T500), as shown in Figure 2. In each array designation, the letter indicates the geometry, C for circular and T for triangular, while the number represents the approximate array size in meters: the diameter for circular arrays and the longest leg for the triangular array. The C50 array consisted of eight seismometers, whereas the C300, T500, and C2150 arrays each consisted of ten. The C300 and T500 arrays shared four receiver locations. The arrays were deployed sequentially rather than concurrently, with the same seismometers relocated for each configuration. MASW testing and deployment of the C50, C300, and T500 arrays were completed during the first field campaign.

Preliminary inversion results showed considerable uncertainty in the shear-wave velocity ($V_S$) of the bedrock. To reduce this uncertainty, the larger C2150 array was deployed during the second campaign. This array extended the measured dispersion data to lower frequencies, thereby providing improved constraints on the depth and $V_S$ of the bedrock.

Each MAM array recorded continuously at a sampling frequency of 250 Hz. Recording durations were approximately 1 hour for C50, 1.5 hours for C300, 3 hours for C2150, and 9.5 hours for T500, which was left recording overnight. Ambient-noise records from all MAM stations were also used in the HVSR analysis. In addition, a seismometer was deployed near the strong-motion station N4.N62A location to record ambient noise for 2 hours and 40 minutes.

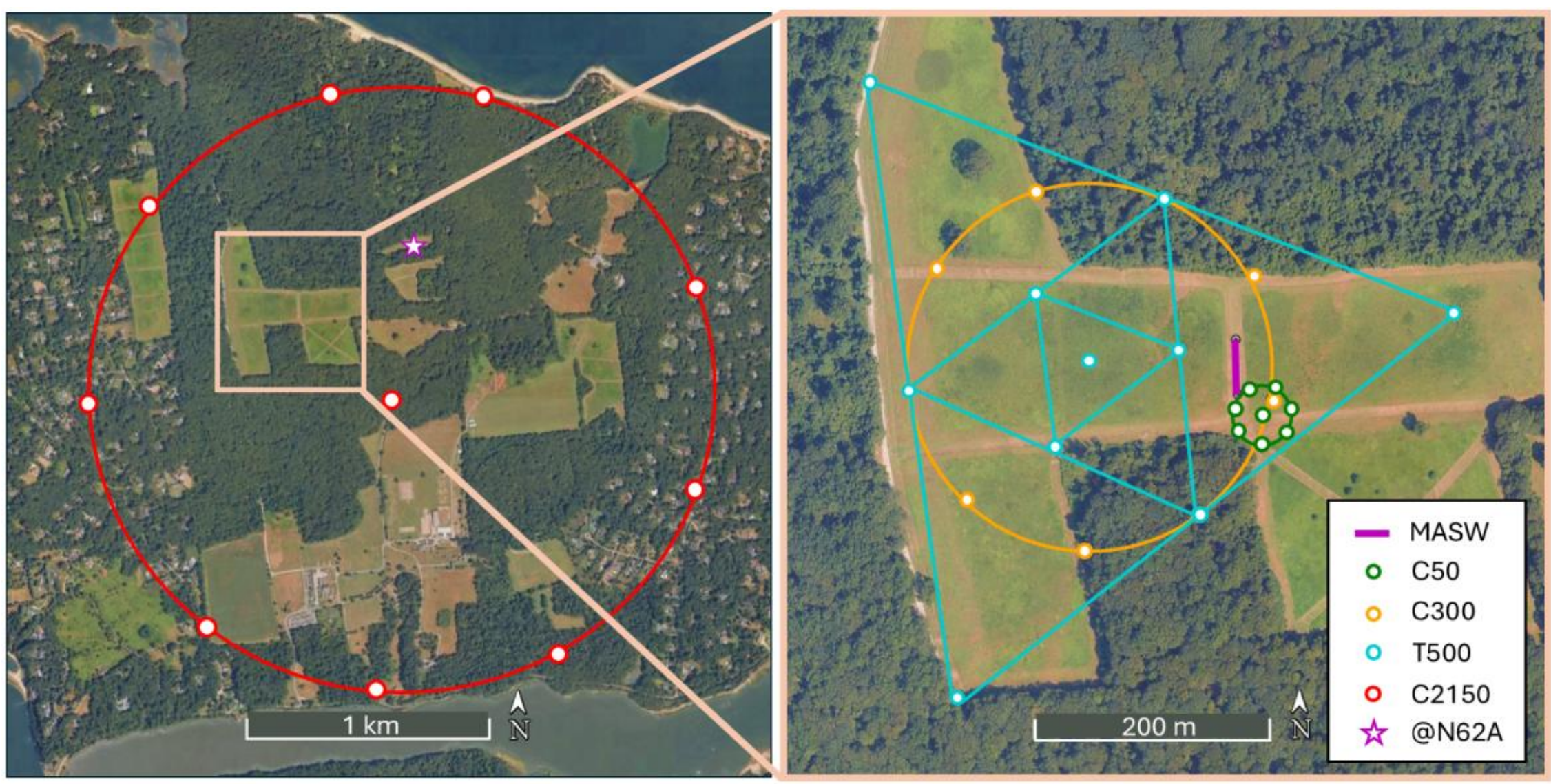

*Figure 2. Caumsett State Historic Park site and array layout shown on Google Earth imagery. Left: the 2,150 m circular MAM array (C2150), shown in red. Right: the MASW survey line, shown in purple; the 50 m circular MAM array (C50), shown in green; the 300 m circular MAM array (C300), shown in orange; and the 500 m concentric triangular MAM array (T500), shown in cyan.*

***Data Processing***

Rayleigh- and Love-wave phase-velocity dispersion data were processed from the MASW records using frequency-domain beamforming with cylindrical-wave steering (Zywicki, 1999), as implemented in the open-source surface-wave processing package swprocess (Vantassel, 2021). This approach was combined with the multiple-source-offset technique to identify near-field contamination and quantify dispersion uncertainty (Cox and Wood, 2011). MASW dispersion data influenced by near-field effects or significant off-line noise were trimmed. Dispersion data were extracted from the MAM records using three-component beamforming (Wathelet et al., 2018), as coded in the open-source software package Geopsy (Wathelet et al., 2020). Spurious dispersion data resulting from high-amplitude near-field or incoherent noise were manually removed before the dispersion statistics were calculated. The active- and passive-source datasets were then combined into a single experimental dispersion dataset. The mean and plus or minus one standard deviation of the Rayleigh- and Love-wave dispersion data were calculated following Vantassel and Cox (2022). The combined dispersion dataset and theoretical resolution limits are shown in Figure 3. Dispersion estimates from the independent arrays overlap and form a continuous trend, supporting the consistency of the combined dataset. The dataset spans frequencies from approximately 0.2 to 50 Hz, corresponding to wavelengths of approximately 4 to 10,000 m. The Rayleigh-wave MASW data constrain wavelengths of roughly 4 to 45 m, while the four MAM arrays progressively extend the dataset to longer wavelengths and lower frequencies. Based on the geometry of the largest array (C2150), the maximum reliably resolvable wavelength, corresponding to the $k_{\min}/2$ resolution criterion of Wathelet et al. (2008), is approximately 6,280 m. Dispersion estimates at longer wavelengths may be affected by the limited array aperture and therefore have greater uncertainty.

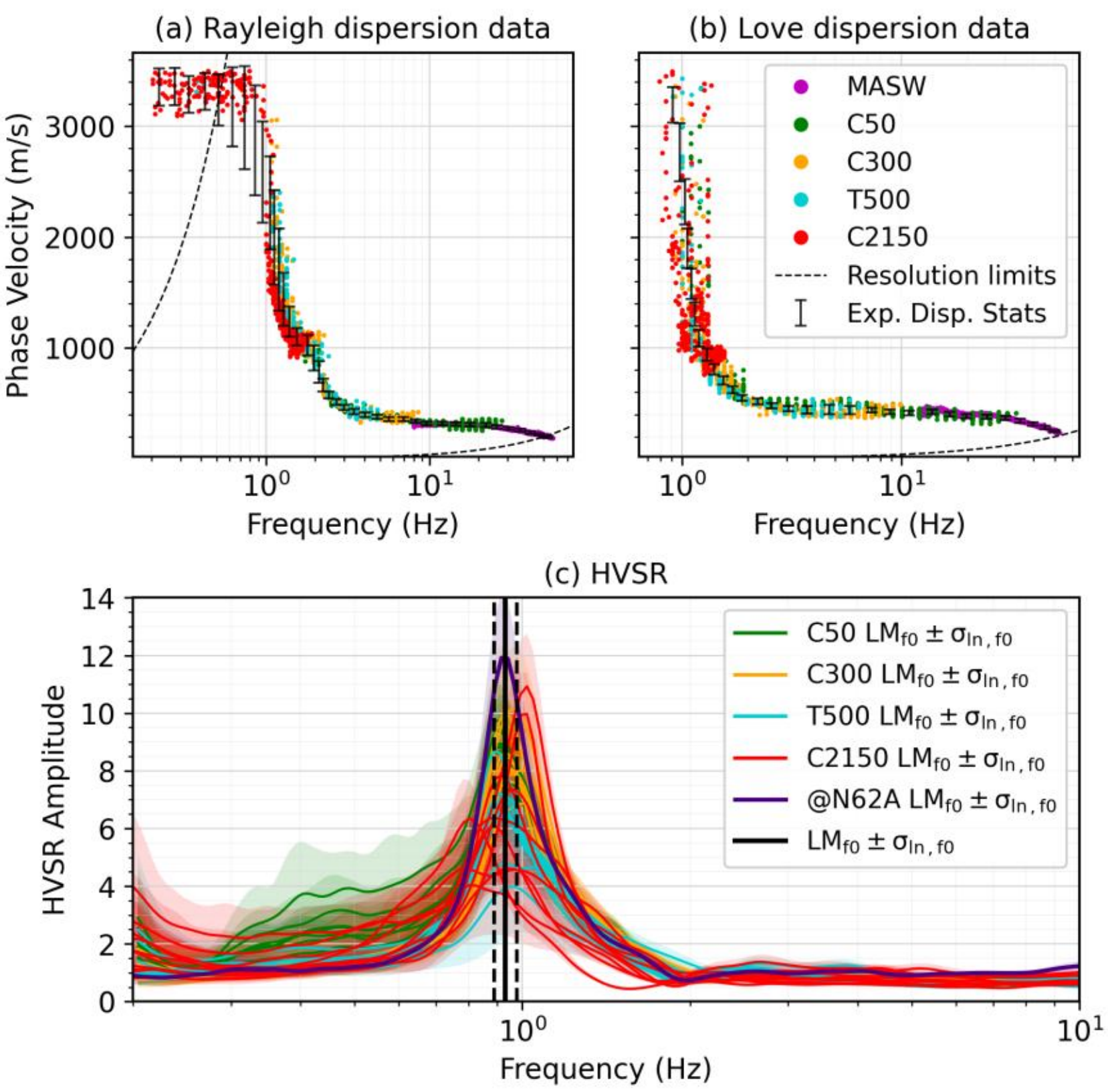

*Figure 3. Experimental surface-wave dispersion and HVSR results for Caumsett State Historic Park. (a) Rayleigh-wave and (b) Love-wave phase-velocity dispersion data obtained from MASW and the C50, C300, T500, and C2150 MAM arrays. Colors identify the testing configuration, and the black dashed curves indicate the theoretical array-resolution limits. (c) Single-station HVSR curves for the MAM stations and the station near strong-motion station N4.N62A, colored by array. The black vertical lines indicate the spatially weighted lognormal median peak frequency $LM_{f0}$ = 0.93 Hz, and the corresponding uncertainty bounds defined by a natural-log standard deviation ($\sigma_{ln,f0}$) equal to 0.05, calculated using stations from the C50, C300, and T500 arrays.*

HVSR curves were computed for all stations used in MAM testing, as well as for the station placed near the strong-motion station N4.N62A, using the frequency-domain window-rejection algorithm (Cox et al., 2020) and the open-source Python package hvsrpy (Vantassel, 2020). The resulting HVSR curves are shown in Figure 3c. When a well-defined peak is present in an HVSR curve, its lowest peak frequency, $f_{0_{HVSR}}$, can provide an estimate of the site fundamental shear-wave resonance frequency, $f_{0_S}$ (Lermo and Chávez-García, 1993; Lachet and Bard, 1994; SESAME, 2004) and/or the lowest-frequency peak associated with fundamental-mode Rayleigh-wave ellipticity, $f_{0_R}$ (Malischewsky and Scherbaum, 2004; Poggi and Fäh, 2010). Following Cheng et al. (2021), Voronoi tessellation was used to decluster and spatially weight the distributed HVSR measurements, thereby obtaining spatially representative site-wide estimates of the lognormal median frequency, $LM_{f0}$, and its associated natural-log standard deviation, $\sigma_{ln,f0}$. For the Caumsett site, the spatially averaged statistics calculated using stations from the C50, C300, and T500 MAM arrays were $LM_{f0}$ = 0.93 Hz and $\sigma_{ln,f0}$ = 0.05, as shown in Figure 3c. The spatially averaged median frequency was nearly identical to the $f_{0_{HVSR}}$ value obtained at the station near N4.N62A. The consistency of the $f_{0_{HVSR}}$ estimates across these arrays indicates limited lateral variability in the subsurface structure over the corresponding spatial extent, supporting the one-dimensional (1D) assumption for site-response analysis. The well-defined peak suggests a strong impedance contrast at depth. When stations from the C2150 array were also included in the spatial weighting, thereby extending the analysis over a broader area, $LM_{f0}$ decreased slightly to 0.91 Hz, while $\sigma_{ln,f0}$ increased to 0.09. These $f_{0_{HVSR}}$ measurements provide an independent constraint for evaluating whether candidate $V_S$ profiles reproduce the observed site resonance.

***Inversion***

The objective of surface-wave inversion is to determine layered-earth models that reproduce the experimental observations through forward modeling. Because the inverse problem is nonunique, it is best approached systematically by evaluating multiple representations of the subsurface. Two parameterization approaches, which control how the subsurface is divided into layers and how layer thicknesses vary with depth, were considered: layering by ratio (LR) and layering by number (LN) (Cox and Teague, 2016; Vantassel and Cox, 2021). For each approach, trial models spanning a range of layer numbers and depths were evaluated. The inversions were performed using the neighborhood algorithm implemented in Geopsy (Sambridge, 1999; Wathelet et al., 2004).

The inversion misfit function combined a Rayleigh-wave dispersion term and a Rayleigh-wave ellipticity-peak term, with the latter constrained by the measured HVSR peak from the C50, C300, and T500 arrays. Love-wave dispersion data were withheld from the inversion and subsequently used as an independent check of whether the resulting $V_S$ profiles could also reproduce the Love-wave observations. Numerous parameterizations and modal interpretations were evaluated, including models containing velocity reversals and interpretations in which portions of the experimental Rayleigh-wave dispersion data were assigned to higher modes. More than 60,000 trial models were evaluated for each parameterization. The results presented in Figure 4 comprise the 100 lowest-misfit profiles from each of the eight best-performing parameterizations, yielding a total of 800 profiles that collectively represent parameterization and inversion uncertainty. The full-depth profiles are shown in Figure 4c, while the median profile from each parameterization is shown over the upper 800 m in Figure 4d. Additional results for the eight median profiles and their numerical values are provided in Figure S2 and Table S1. The profiles are interpreted with the greatest confidence to a depth of approximately 2,500 m, estimated by dividing the maximum reliably resolved wavelength, $\lambda_{res}$, by 2.5 (Foti et al., 2015; Garofalo et al., 2016). The $V_S$ values below this depth are presented only as qualitative guidance.

All eight parameterizations produced Rayleigh theoretical dispersion curves that fit the experimental data within their uncertainty bounds, with misfit values ranging from 0.28 to 0.59 (Figure 4a). The selected $V_S$ profiles reproduced the general trend of the withheld Love-wave dispersion data, providing an independent check on the inverted velocity structure (Figure S1b). The corresponding theoretical Rayleigh-wave ellipticity curves also exhibit primary peaks that cluster near the measured $LM_{f_0}$ of 0.93 Hz, demonstrating consistency between the dispersion and HVSR constraints (Figure 4b). The fundamental frequencies of the associated 1D shear-wave transfer functions similarly agree with the measured $LM_{f0}$ (Figures 5 and S1d). The mean $V_{S30}$ across all 800 profiles is 349 m/s, with a standard deviation of 10 m/s (Figure S4). Within individual layers, the variability in $V_S$ is generally between 0.05 and 0.10 in natural-log units; the depth-dependent standard deviation across all 800 profiles is shown in Figure S3. Despite differences among the parameterizations, all models resolve the same first-order structure: sediments that gradually stiffen with depth overlie a pronounced impedance contrast, below which $V_S$ increases to values consistent with crystalline bedrock (Figures 4c and 4d).

A nearby water well documented in the NYSDEC Well Log Database (Well S136718) provides an independent check on the inferred subsurface structure. The log, which was withheld from the inversion, places bedrock at a depth of approximately 177 m. The depths of the inferred impedance contrast are broadly consistent with this value, supporting the deep site-characterization results.

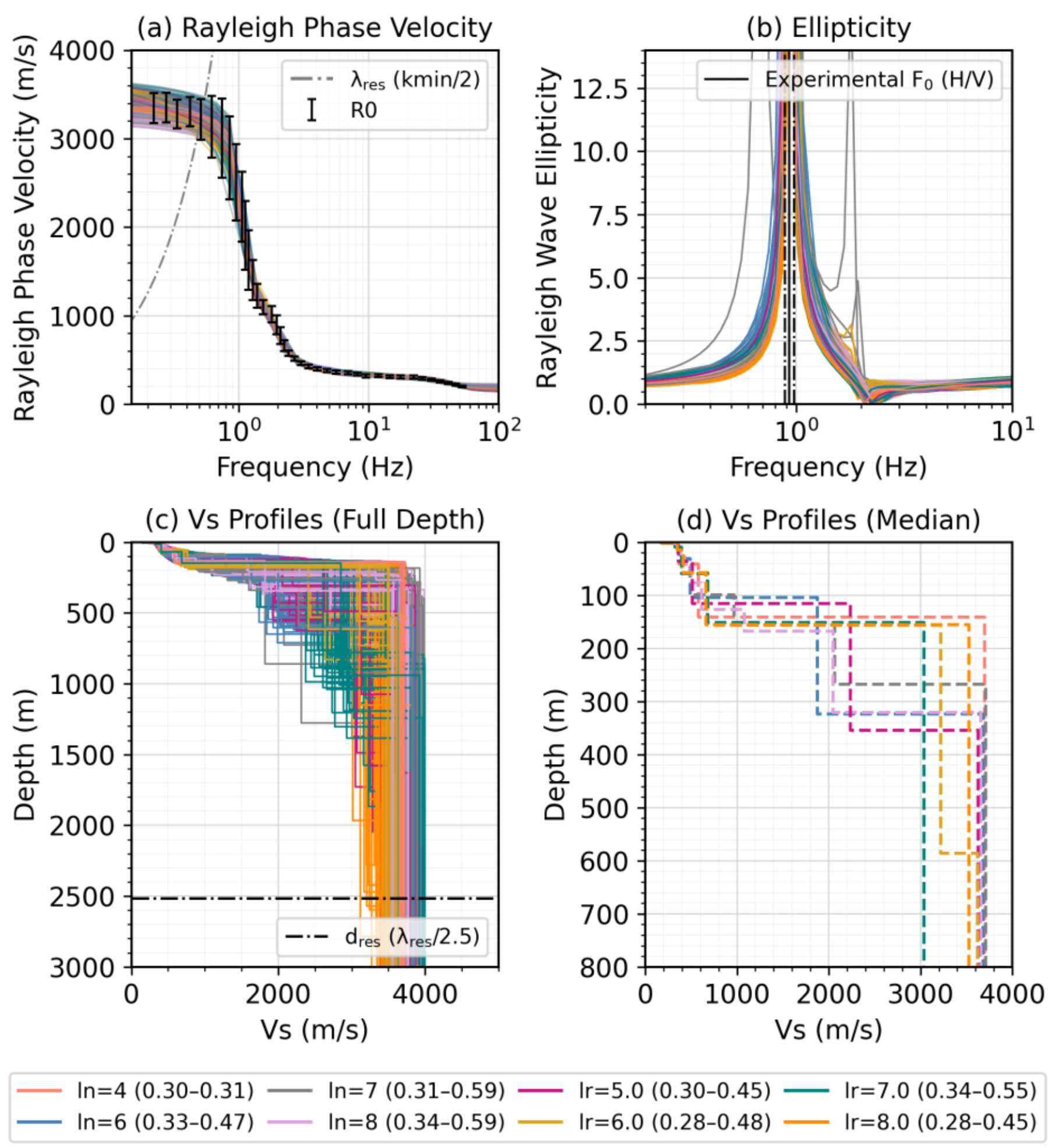

*Figure 4. Surface-wave inversion results for the eight best-performing parameterizations. (a) Experimental fundamental-mode Rayleigh-wave phase-velocity dispersion statistics (R0; mean and plus or minus one standard deviation) and theoretical dispersion curves for the 100 lowest-misfit models from each parameterization. The gray dash-dotted curve indicates the* $k_{min}/2$ *array-resolution limit. (b) Corresponding fundamental-mode Rayleigh-wave ellipticity curves. The black vertical lines indicate the measured HVSR lognormal median peak frequency,* $LM_{f0} = 0.93$ *Hz, and its plus or minus one natural-log standard deviation bounds. (c) Full-depth* $V_S$ *profiles for all 800 selected models. The horizontal dash-dotted line indicates the approximate resolution depth,* $d_{res} = \lambda_{res}/2.5 \approx 2{,}500$ *m; profiles below this depth are shown for guidance only. (d) Median* $V_S$ *profile from each parameterization over the upper 800 m. Colors identify the parameterizations (ln: layering by number; lr: layering by ratio), and the values in parentheses in the legend indicate the range of misfit values among the 100 selected models for each parameterization.*

### 3.2 Discussion: Sediment Resonance at Caumsett

To evaluate sediment resonance at Caumsett using the site characterization data, Figure 5 compares four representations of linear site amplification: (1) 1D linear theoretical transfer functions calculated from the eight median $V_S$ profiles shown in Figure 4d, (2) the ergodic CEUS NGA-East $V_{S30}$-based linear amplification model of Stewart et al. (2020), (3) the ergodic CEUS linear amplification model of Meyer et al. (2026) based on geospatial predictor variables (adjusted physiographic province and sediment thickness), and (4) observed amplifications from S2S residuals from two recent earthquakes recorded at this station. The 1D linear theoretical transfer functions are computed from the $V_S$ profiles using the Thomson-Haskell matrix method (Haskell, 1953; Thomson, 1950); the material density and quality factor of each layer are estimated from $V_S$ using Boore (2016) and Campbell (2009), respectively. The Stewart et al. (2020) ergodic model uses $V_{S30}$ = 350 m/s (consistent with the mean $V_{S30}$ from the profiles in Figure 4d), and the Meyer et al. (2026) ergodic model uses a sediment thickness of 177 m (consistent with Well S136718) and the Embayed physiographic province. Note that the Meyer et al. (2026) linear amplification model is usually paired with sediment thicknesses from Boyd et al. (2024) in the coastal plain and Pelletier et al. (2016) elsewhere. At this site, sediment thickness from the borehole log was used because it provides a direct, site-specific constraint. However, the different sediment-thickness estimates yielded similar amplification values. The observed S2S-based amplification is computed as $\delta S2S + F_{lin}$. Here $\delta S2S$ quantifies the repeatable over- or underprediction of recorded ground motions at a given station by the NGA-East GMM, and $F_{lin}$ is the NGA-East linear amplification from Stewart et al. (2020).

Across all $V_S$ parameterizations, the 1D theoretical transfer functions consistently reproduce two principal amplification features observed in the S2S-based amplification: the fundamental resonance frequency, $f_0$, near approximately 1.0 Hz, and the first higher-mode resonance frequency near 2.5 Hz. This $f_0$ range is consistent with the measured HVSR peak frequency, $LM_{f_0} = 0.93$ Hz, used to constrain the inversion, as well as the 1-Hz frequency at which N4.N62A recorded a large positive pseudospectral-acceleration residual during the 2024 Tewksbury earthquake. Beyond the first higher mode, however, the predicted frequencies and amplitudes of the higher-mode resonances vary more among the alternative $V_S$ parameterizations.

The two CEUS ergodic linear amplification models provide smoother representations of the Caumsett response. The NGA-East $V_{S30}$-based model of Stewart et al. (2020) predicts broad amplification but does not reproduce the distinct resonance features observed in the 1D theoretical transfer functions and S2S-based amplification. By incorporating the geospatial parameters of sediment thickness and adjusted physiographic province, the Meyer et al. (2026) model is in stronger agreement with both the theoretical transfer functions and the S2S-based amplifications. The Meyer et al. (2026) model more closely captures the frequency range of elevated amplification around $f_0$, although its response remains too smooth to fully reproduce the pronounced fundamental and first higher-mode resonances. These comparisons indicate that smooth, ergodic linear site-amplification models parameterized by sediment thickness improve the regional representation of low-frequency response but do not fully capture site-specific modal behavior at Caumsett.

Because the theoretical transfer functions, ergodic models, and S2S-based amplifications represent different measures of site response and use different reference conditions, Figure 5 is interpreted primarily in terms of resonance frequency and overall response shape rather than exact amplification amplitude. The transfer functions are Fourier spectral ratios calculated from the inverted $V_S$ profiles, whereas the ergodic models describe acceleration response-spectral amplification relative to a hard-rock reference condition. Fourier and response-spectral ratios can identify similar resonant frequencies under linear conditions (Tao and Rathje, 2019), but response-spectral amplification also depends on the input ground motion (Bora et al., 2016; Dobry et al., 2000; Zhao et al., 2009). The strong agreement of the S2S-based amplifications with the transfer functions at $f_0$ and the first higher-mode resonance frequency, together with the large positive 1-Hz residual recorded during the 2024 Tewksbury earthquake, provides independent evidence that resonance of the deep sedimentary column is the dominant amplification mechanism at N4.N62A. This conclusion is largely in agreement with patterns observed at a deep sediment site in New York City evaluated by Kaklamanos et al. (2026).

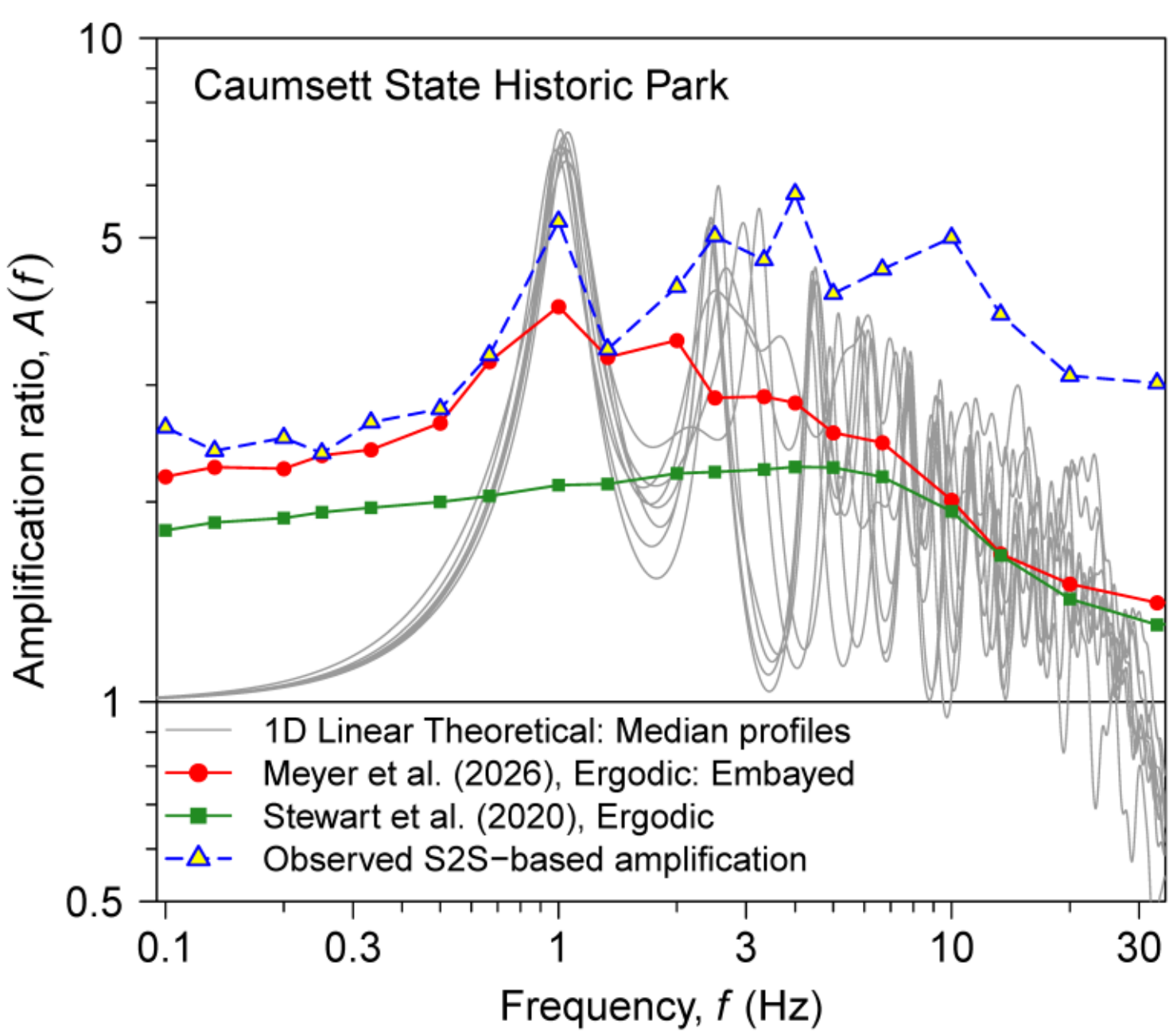

***Figure 5.*** *Comparisons of observed and predicted amplification estimates for Caumsett State Historic Park. One-dimensional linear transfer functions calculated from the eight median* $V_S$ *profiles obtained from the alternative surface-wave inversion parameterizations (Figure 4d) are compared with the ergodic linear site-amplification models of Stewart et al. (2020) and Meyer et al. (2026), and observed amplifications computed from site-to-site residuals from two earthquakes recorded at N4.N62A.*

## 4. LAMONT-DOHERTY EARTH OBSERVATORY

### 4.1 Site Description and Testing Methods

Strong-motion station LD.PAL is located at the Lamont-Doherty Earth Observatory in southeastern New York, approximately 170 m west of the crest of the Hudson River Palisades. The Palisades form a prominent bedrock ridge along the western shore of the Hudson River and are underlain primarily by Jurassic diabase associated with Early Mesozoic rifting (Soren, 1988;

Puffer et al., 2009). The ridge extends along the Hudson River from approximately the George Washington Bridge to Haverstraw, New York (Puffer et al., 2009), a distance of approximately 40 km. Near the observatory, the ridge rises approximately 115 m above the Hudson River, with a steep eastern flank descending toward the river and a more gradual western slope extending inland, as shown in Figure 6. The main ridge segment south of the site trends approximately N18°E, whereas near the observatory the ridge crest broadens and becomes more irregular, with local crest segments oriented in different directions. Across the site, the diabase is either exposed at the ground surface or overlain by a thin sediment cover that varies laterally in thickness.

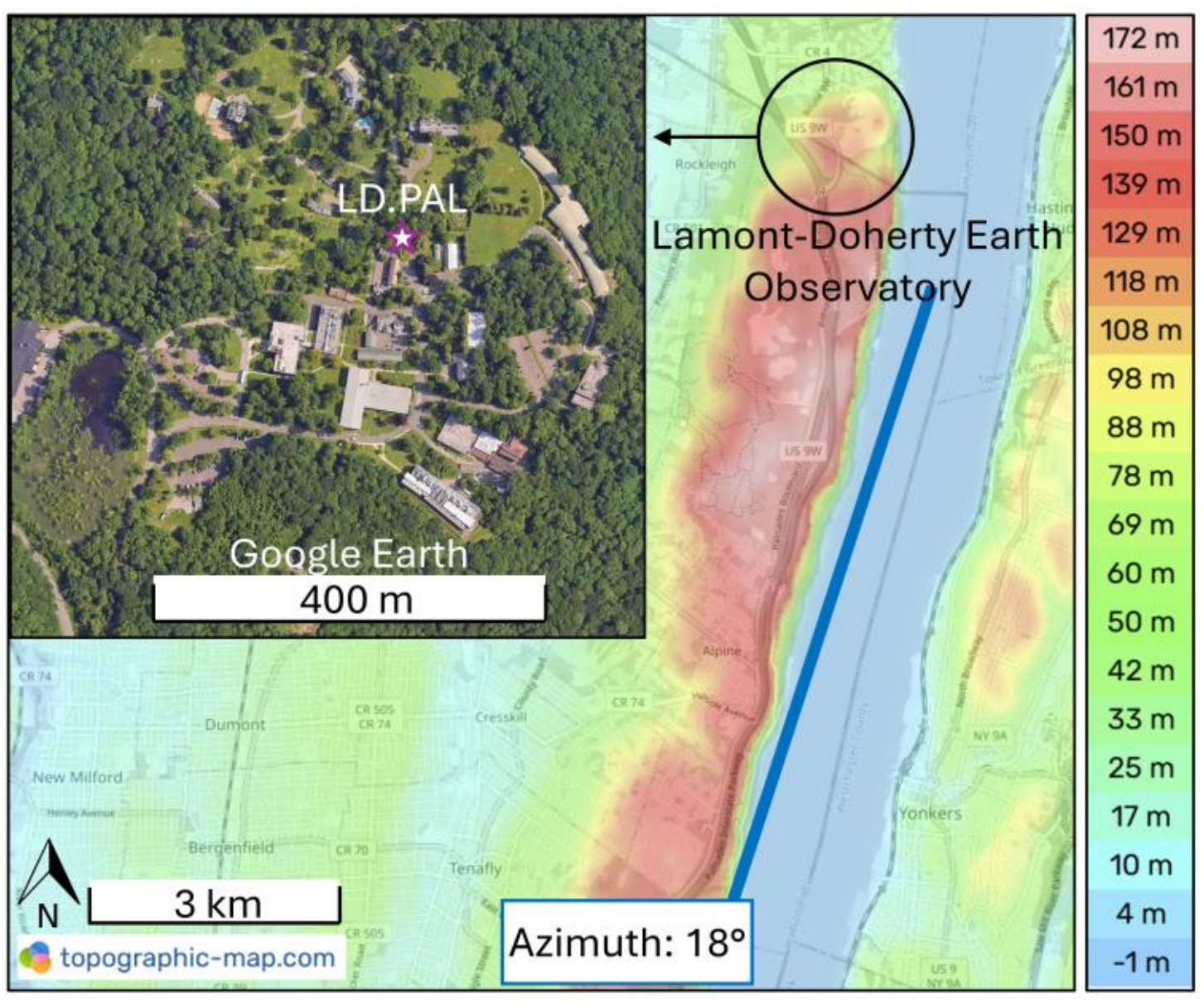

*Figure 6. Topography of the Hudson River Palisades near the Lamont-Doherty Earth Observatory. The elevation map illustrates the steep eastern flank of the ridge along the Hudson River and the more gradual terrain extending westward. The blue line marks the main ridge segment south of the site, which trends approximately N18°E. The black circle identifies the observatory area, shown in greater detail in the Google Earth inset. Elevations are shown in meters. Base imagery from Google Earth and topographic data from topographic-map.com.*

The field investigation combined surface-wave testing with two deployments designed to evaluate topographic amplification. Surface-wave testing was conducted to characterize the $V_S$ structure at the site and assess whether 1D site response could account for the observed amplification. The two topographic-amplification deployments were used to investigate how amplification varied with frequency, direction, and spatial position across the site, and they were carried out sequentially so that the first could inform the design of the second. The surface-wave and topographic-amplification testing procedures are described below.

*Surface-Wave Testing*

Surface-wave testing consisted of MASW and MAM, with the MAM measurements conducted using circular arrays approximately 50 and 300 m in diameter. The two MAM arrays recorded ambient vibrations for approximately 1 and 2 hours, respectively. Data acquisition, dispersion processing, and inversion generally followed the procedures described for the Caumsett site in Section 3. Additional details on local near-surface variability and the experimental dispersion target are provided in the Supplementary Material (Section S2, Figures S5 and S6). The resulting profiles indicate approximately 2 m of sediment at the MASW testing location, with $V_S$ increasing rapidly with depth to values around 3,150 m/s (Figures S7 and S8; Table S2). The corresponding mean $V_{S30}$ is 1,630 m/s, with a standard deviation of 20 m/s, consistent with hard-rock site conditions. HVSR curves were calculated for each MAM station and for the stations of the two topographic-amplification deployments. Most HVSR curves exhibited modest peaks well above the 1-Hz frequency at which the large positive Tewksbury residual was observed. Peak frequencies varied considerably across the site, likely reflecting lateral changes in the thickness of the surficial cover (Figure S5).

*Topographic amplification testing*

The topographic-amplification investigation was designed to quantify how ground-motion amplification varies with frequency, direction, and location across the site using ambient-noise recordings. Ambient-noise measurements have been shown to produce amplification patterns consistent with those obtained from earthquake recordings, supporting their use to investigate small-strain topographic amplification (Cannon, 2024; Rischette et al., 2026). Two arrays were deployed for this purpose, each consisting of ten three-component broadband seismometers (SmartSolo IGU-BD3C-5). Sensor locations were designated St-01 through St-10 within each deployment. The first deployment, LamTA01, spanned inland locations west of the crest and the crest itself, with an additional sensor positioned at the base of the eastern slope near the Hudson River. The first deployment also included one sensor collocated with strong-motion station

LD.PAL (St-10), as shown in Figure 7. LamTA01 recorded approximately 13 hours of continuous ambient noise overnight from July 30 to 31, 2025.

The second deployment, LamTA02, was informed by the amplification pattern observed during LamTA01. The far-field reference location (St-01) and the LD.PAL location (St-10) were reoccupied to test repeatability under different ambient-noise and sensor-coupling conditions, and their identifiers are retained across both deployments. The remaining eight sensors were repositioned across the crest and eastern slope, as shown in Figure 8. LamTA02 recorded approximately 18 hours of continuous ambient noise from November 30 to December 1, 2025. All sensors were buried and oriented toward magnetic north, and weighted buckets were placed over them to reduce wind noise. Figures 7 and 8 also present the directional spectral-ratio results; the processing and interpretation of the polar plots are described below.

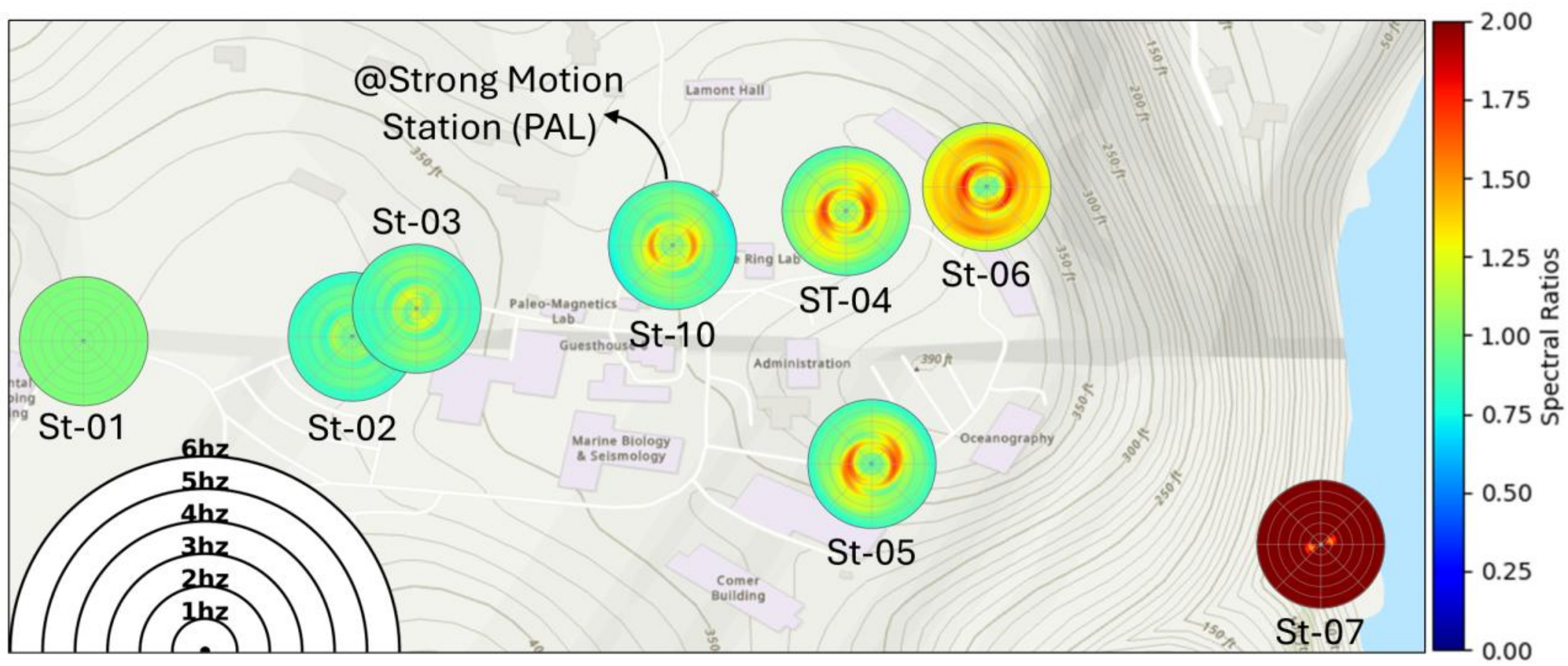

*Figure 7. Sensor layout and directional spectral-ratio results for the first topographic-amplification deployment (LamTA01) at the Lamont-Doherty Earth Observatory. Polar plots are centered at the locations of stations St-01 through St-07 and St-10, with St-10 collocated with strong-motion station LD.PAL. Color indicates spectral-ratio amplitude and radial distance represents frequency from 1 to 6 Hz. Background contours show the local topography.*

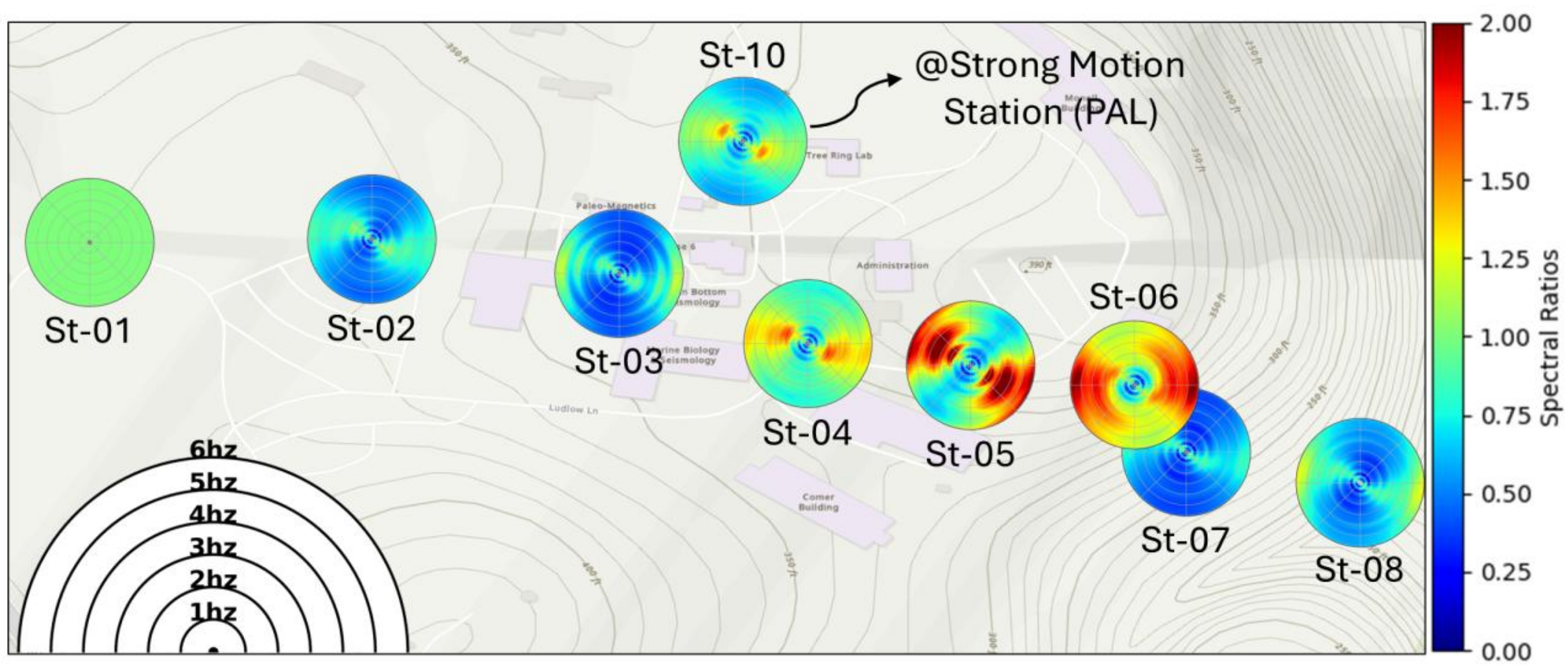

*Figure 8. Sensor layout and directional spectral-ratio results for the second topographic-amplification deployment (LamTA02). Polar plots are centered at stations St-01 through St-08 and St-10, with St-10 collocated with strong-motion station LD.PAL. Color indicates spectral-ratio amplitude and radial distance represents frequency from 1 to 6 Hz. Background contours show the local topography.*

To process the ambient-noise data and quantify topographic amplification, the ground-motion spectrum at each station was divided by that from a reference location sufficiently far from the crest to be minimally affected by the topography. Amplification was quantified using two spectral-ratio methods: the standard spectral ratio (SSR; Borcherdt, 1970) and the median reference method (MRM; Wilson and Pavlis, 2000; Stolte et al., 2017). Both methods divide the Fourier amplitude spectrum (FAS) of a given station by a reference spectrum; the SSR uses a single reference station, whereas the MRM uses the lognormal median spectrum from a group of stations. Both methods were applied to evaluate the sensitivity of the results to reference selection (Cannon, 2024). For consistency between the two deployments, only the SSR results are presented in this paper, using far-field station St-01 as the common reference.

Each record was bandpass filtered between 0.1 and 40 Hz using a fifth-order Butterworth filter and divided into 5-minute time windows, which are long enough to capture approximately 150 cycles of motion at 0.5 Hz (Stolte et al., 2017). Each window was linearly detrended and tapered using a Tukey window with a length equal to 5% of the window length applied to each end. The sensor orientations were corrected for the magnetic declination at the site, approximately −12.6° during both deployments, so that all azimuths were referenced to true north. The two orthogonal horizontal components were then rotated to resolve ground motion along azimuths spanning 0 to 180 degrees in 3-degree increments; because motion along opposite azimuths is identical, this yields 60 unique median spectral-ratio curves per station. For each window and azimuth, the FAS was computed and smoothed using the Konno and Ohmachi (1998) method with a bandwidth coefficient of 40. Spectral ratios relative to St-01 were calculated window by window at frequencies between 0.2 and 6 Hz and reduced to a single median spectral-ratio curve per azimuth by taking the median across all windows, yielding 60 curves for each station.

The directional spectral-ratio results can be displayed either as spectral-ratio curves or as polar plots, as illustrated in Figure 9. The two panels contain the same values but organize them differently. In Figure 9a, each curve shows spectral ratio as a function of frequency for one azimuth, with line color indicating the azimuth range. For example, the red curve corresponding to an azimuth of 113° reaches a spectral ratio of 1.8 at 2.2 Hz, indicating that, during LamTA01, the Fourier amplitude at St-06 exceeded that at reference station St-01 by a median factor of 1.8 for horizontal motion resolved along the 113° azimuth at 2.2 Hz. Figure 9b presents the same data as a polar plot, in which angular position indicates azimuth measured from true north, radial distance indicates frequency increasing outward from the center, and color indicates spectral ratio. That same peak appears in Figure 9b at the two white markers. The markers lie along the white dashed line joining the 113° and 293° directions at the radial distance corresponding to 2.2 Hz, while the red shading at those locations indicates a spectral ratio of 1.8. Two markers are shown because horizontal motions resolved along opposite directions on the same axis are identical. Because polar plots make the dominant frequencies and directions of amplification easier to visualize for multiple stations in the same figure, they are used to present the spatial results in Figures 7 and 8.

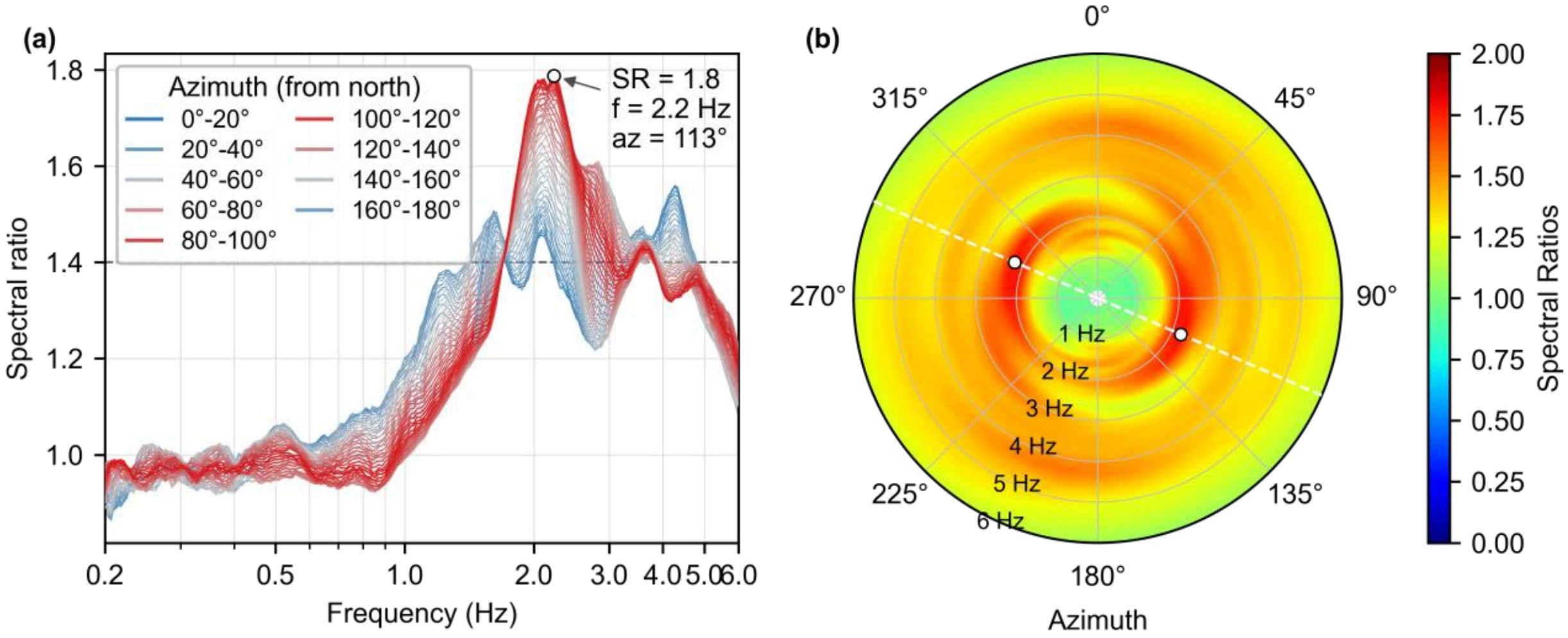

*Figure 9. Equivalent representations of the directional spectral-ratio results for LamTA01 St-06. (a) Median spectral-ratio curves plotted as a function of frequency for each analyzed azimuth, with line color indicating azimuth range. The marked point identifies the highest amplification at this station, with a spectral ratio of 1.8 at 2.2 Hz along an azimuth of 113°. (b) Polar representation of the same data, in which angular position indicates azimuth measured from true north, radial distance indicates frequency, and color indicates spectral-ratio amplitude. The dashed line and white markers identify the same direction and frequency of maximum amplification.*

The measured spectral-ratio results were evaluated against four criteria for identifying topographic amplification (Ashford et al., 1997; Massa et al., 2010; Cannon, 2024). First, the spectral-ratio curves exhibit clear amplification, with peak amplitudes greater than approximately 1.4. Second, the strongest amplification is oriented approximately perpendicular to the elongation of the ridge. Third, amplification at crest stations exceeds that at slope and behind-crest stations. Fourth, the amplification occurs near the resonance frequency estimated from the ridge geometry, $f \approx V_S/(5H)$, where $V_S$ is the shear-wave velocity of the ridge material and $H$ is the ridge height (Ashford et al., 1997), or at least within the range of 1 to 6 Hz commonly observed in experimental studies (Cannon, 2024).

Figures 7 and 8 show that the results from both deployments generally satisfy the four criteria for topographic amplification. During LamTA01, the strongest amplification occurred at St-05 and St-06, which were located near local ridge crests and exhibited peak spectral ratios greater than 1.4. These stations display the darkest colors in their polar plots. Amplification generally decreased at stations farther inland, including St-04, St-10, and St-03, as indicated by the progressively lighter colors. A similar pattern was observed during LamTA02. Peak spectral ratios again exceeded 1.4 at crest stations St-05 and St-06, while smaller ratios were observed downslope at St-07 and inland at St-04, St-10, and St-03. Nearly all stations also showed preferential directions of amplification, which were broadly consistent with directions perpendicular to nearby crest segments. Amplification was observed from approximately 2 Hz to the 6 Hz upper limit of the analyzed frequency range. This band broadly agrees with the estimate $f \approx V_S/(5H)$ and falls within the 1 to 6 Hz range commonly observed in experimental studies. Using a ridge height of $H = 115$ m and $V_S$ values ranging from the measured $V_{S30}$ of 1,630 m/s to the deeper-rock velocity of approximately 3,150 m/s, yields an estimated frequency range of approximately 2.8 to 5.5 Hz (Ashford et al., 1997). Together, these observations show that amplification exceeds the 1.4 threshold, is strongest near the crests, decreases away from them, occurs within the frequency range associated with topographic amplification, and is concentrated along preferential directions. The results are therefore consistent with topographic amplification.

St-06, located near the ridge crest during LamTA01, provides a clear example of the relationship between the direction of amplification and the local topography (Figure 9). The main ridge segment south of the site trends approximately N18°E (Figure 6), giving a perpendicular direction near 108°, which agrees closely with the 113° azimuth of the strongest response at this station. This agreement indicates that the strongest amplification is approximately perpendicular to the main ridge. The polar plot also shows additional amplification peaks at other frequencies and directions, which may reflect the influence of local ridge segments with different orientations.

St-07 during LamTA01 requires separate consideration because its location beside the Hudson River exposed it to strong local noise. The station was positioned at the toe of the eastern slope, while the St-01 reference station was located far inland and at a much higher elevation. The two stations therefore likely did not record the same river-generated wavefield, causing the spectral ratios at St-07 to be strongly influenced by local noise. Consequently, the broadly elevated spectral ratios at St-07 should not be interpreted directly as topographic amplification. However, the station still exhibited deamplification in a direction approximately perpendicular to the ridge, which is consistent with a topographic effect and the redistribution of wave energy near the base of the ridge. Other stations that showed substantial local-noise contamination were excluded from the directional amplification analysis, specifically St-08 and St-09 during LamTA01 and St-09 during LamTA02.

To evaluate the repeatability of the measured response, St-01 and St-10 were placed at the same locations during both LamTA01 and LamTA02. St-01 served as the common far-field reference, while St-10 was collocated with strong-motion station LD.PAL. LD.PAL is located approximately 170 m inland from the ridge crest, well within the 4H distance of approximately 460 m over which topographic effects are expected to persist for a ridge height of 115 m (Ashford et al., 1997). Reoccupying both locations allowed the spectral ratio at LD.PAL to be calculated using the same reference location in each deployment. It also provided a check on whether the results were sensitive to differences in sensor installation and ground coupling. Although the measurements were collected four months apart under different seasonal and ambient-noise conditions, they show similar amplification characteristics in terms of frequency, amplitude, and direction (Figure 10). In both deployments, amplification is concentrated near 2 Hz, reaches spectral ratios of approximately 1.6, and is strongest along an approximately east–west direction. This agreement indicates that the site effect observed at LD.PAL is repeatable rather than an artifact of a particular deployment or sensor installation.

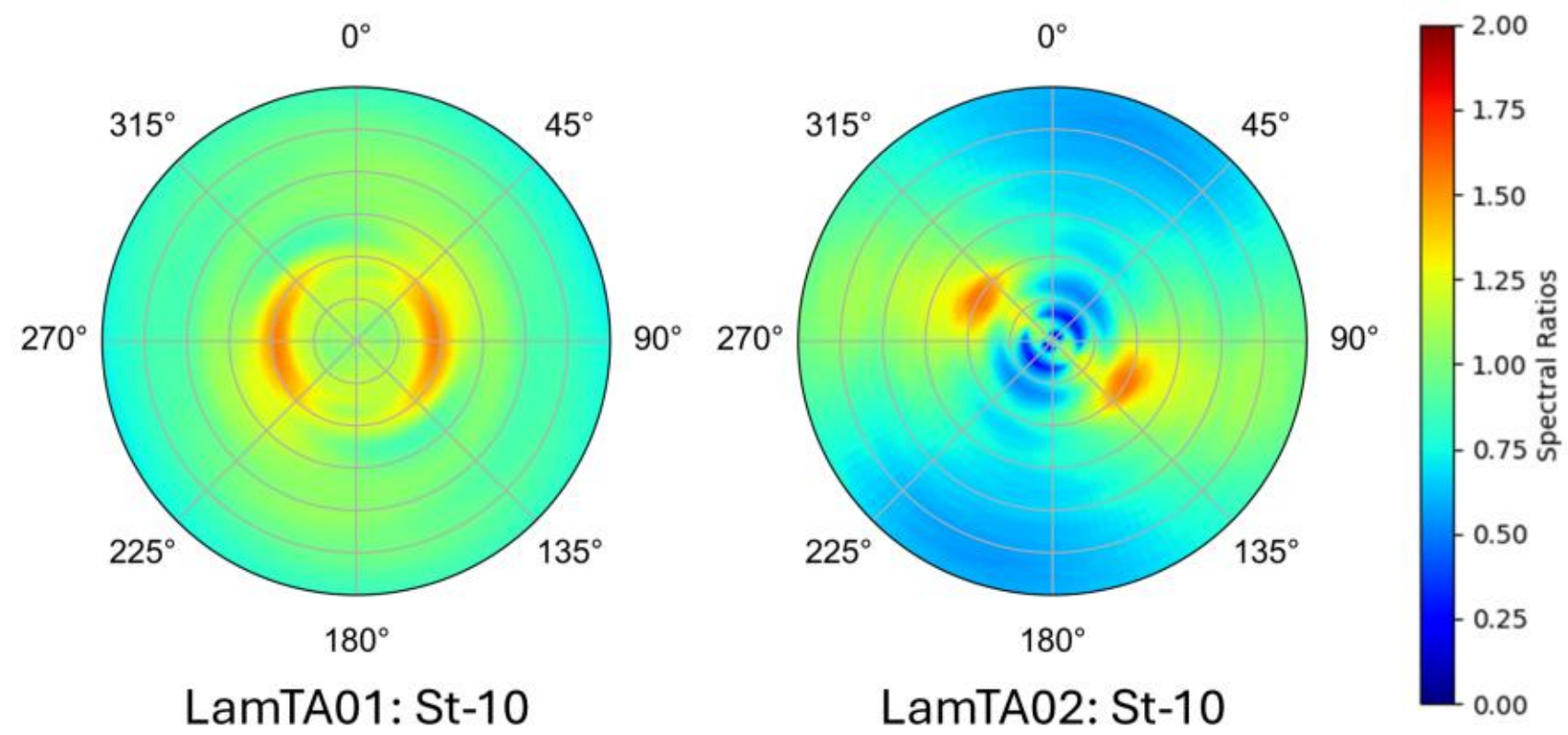

*Figure 10. Comparison of directional spectral-ratio results at strong-motion station LD.PAL during the LamTA01 and LamTA02 deployments. St-10 was collocated with LD.PAL during both deployments, while St-01 was reoccupied as the common far-field reference. Both measurements show similar amplification near 2 Hz, peak spectral ratios of approximately 1.6, and a preferential direction that is approximately east–west.*

### 4.2 Discussion: Topographic Amplification at Lamont-Doherty

To place the Lamont observations in the context of site-effect modeling, Figure 11 compares observations and predictions of amplification at LD.PAL using the same methods employed for the Caumsett site. The theoretical transfer functions calculated using the median $V_S$ profiles from surface-wave testing are compared with the ergodic models of Stewart et al. (2020) and Meyer et al. (2026), as well as the observed S2S amplification derived from ground motions recorded at LD.PAL based on 41 earthquakes. For this comparison, the Stewart et al. (2020) predictions were calculated using $V_{S30}$ = 1,630 m/s. The Meyer et al. (2026) predictions were calculated using a sediment thickness of 2 m for both the New England and Piedmont province assignments because LD.PAL lies near their boundary. As shown in Figure 11, the 1D theoretical transfer functions remain near unity at frequencies below 10 Hz and produce a sharp amplification peak near 30 Hz, with a secondary peak at 80 Hz. This response is associated with the thin sediment cover (~2 m) represented in the inverted profiles. The ergodic linear amplification models produce broader and smoother amplification trends, with the Meyer et al. (2026) predictions varying depending on the province to which the site is assigned. The Meyer et al. (2026) predictions for the Piedmont province are more consistent with the Stewart et al. (2020) predictions and the 1D theoretical transfer functions, but the Meyer et al. (2026) predictions for the New England province are more consistent with the observed S2S-based amplifications, which provide more insight into the behavior at this site.

The observed S2S-based amplification at LD.PAL follows a different pattern of modest amplification between approximately 0.6 and 2 Hz, followed by deamplification at higher frequencies. This low-frequency amplification is not reproduced by either the theoretical transfer functions or the ergodic models. Of the available observations, the S2S-based amplifications agree most closely with the directional amplification near 2 Hz measured at St-10, which was collocated with LD.PAL during both topographic-amplification deployments. Because the S2S residuals are based on recordings from 41 earthquakes, there is a relatively high degree of confidence in the observed amplification pattern from 0.6 to 2 Hz. The agreement in the amplification frequency range provides independent evidence from earthquake recordings that the amplification observed by the temporary arrays is relevant to the permanent strong-motion station, and that topographic effects are likely being represented in the observed ground motions.

The differences among the curves in Figure 11 reflect the processes that each estimate can represent. The 1D theoretical transfer functions describe vertically propagating shear waves through a laterally uniform velocity profile. The ergodic linear amplification models describe average amplification using parameters such as $V_{S30}$, physiographic province, and sediment thickness. Neither approach represents ridge topography or the directional effects documented by the topographic-amplification arrays. Taken together, the observations indicate that the low-frequency amplification at LD.PAL is controlled primarily by the topography rather than by the subsurface velocity structure alone. The thin sediment cover may contribute to amplification at high frequencies, but it cannot explain the observed S2S-based amplification between approximately 0.6 and 2 Hz. A $V_{S30}$ of 1,630 m/s would otherwise characterize LD.PAL as a hard-rock site with limited amplification. These results suggest that representing complex sites such as LD.PAL may require topographic descriptors that account for ridge characteristics.

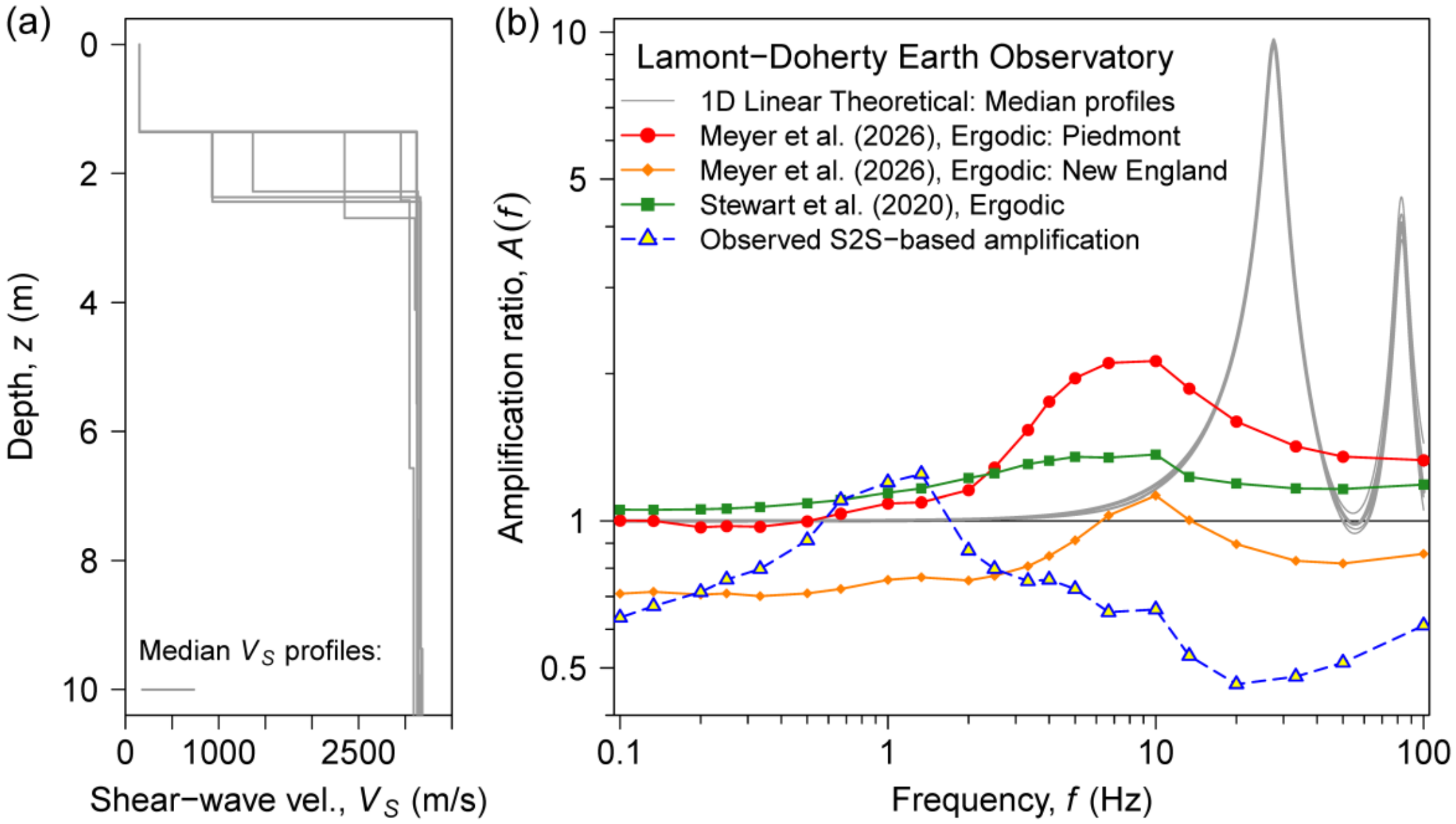

*Figure 11. Comparison of site-specific, ergodic, and observed amplification estimates for strong-motion station LD.PAL. (a) Median $V_S$ profiles for the 100 lowest-misfit models from each of the eight inversion parameterizations. (b) One-dimensional linear theoretical amplification transfer functions calculated from the median $V_S$ profiles, compared with the ergodic amplification models of Stewart et al. (2020) and Meyer et al. (2026) for the Piedmont and New England physiographic provinces, and the observed S2S-based amplification derived from 41 earthquakes recorded at LD.PAL.*

## 5. CONCLUSIONS

This study investigated the physical mechanisms responsible for the amplified ground motions recorded at two strong-motion stations in southeastern New York: N4.N62A at Caumsett State Historic Park and LD.PAL at the Lamont-Doherty Earth Observatory. Both stations exhibited among the largest positive 1-Hz pseudospectral-acceleration residuals in the region during the 2024 $M_w$ 4.8 Tewksbury, New Jersey, earthquake, despite occupying sharply contrasting geologic settings. Detailed geophysical site characterization at both sites combined active-source and ambient-noise surface-wave testing with horizontal-to-vertical spectral ratio (HVSR) measurements. At Lamont-Doherty, two additional ambient-noise arrays were deployed across the site to evaluate the frequency, directionality, and spatial distribution of topographic amplification. The dominant amplification mechanism at each station was then evaluated by comparing 1D theoretical transfer functions calculated from the inverted $V_S$ profiles, empirical ergodic linear amplification models from Stewart et al. (2020) and Meyer et al. (2026), and observed amplifications from site-to-site (S2S) residuals using earthquake recordings.

At Caumsett State Historic Park, the inverted $V_S$ profiles resolved a deep sedimentary column overlying a pronounced impedance contrast. The inferred depth of this contrast is broadly consistent with the bedrock depth of approximately 177 m documented in a nearby water well that was withheld from the inversion. HVSR measurements identified a spatially representative fundamental site frequency of approximately 0.93 Hz, which was used as a constraint during the surface-wave inversion. The transfer functions calculated from the resulting $V_S$ profiles exhibit a fundamental resonance near 1.0 Hz and a first higher-mode resonance near 2.5 Hz. Both resonance features are independently evident in the earthquake-based S2S amplifications. The fundamental-resonance band also includes the 1-Hz frequency at which N4.N62A recorded a large positive residual during the 2024 Tewksbury earthquake. Together, the agreement of the S2S-based amplifications with the transfer functions at the fundamental and first higher-mode resonance frequencies, along with the event-specific 1-Hz residual, supports resonance of the deep sedimentary column as the dominant amplification mechanism at N4.N62A. Comparison with the ergodic models indicates that incorporating sediment thickness and regional geologic context (e.g., Meyer et al., 2026) improves the representation of the broad low-frequency amplification relative to $V_{S30}$-based linear amplification models in the CEUS.

Lamont-Doherty Earth Observatory is a hard-rock site with a thin, laterally variable sediment cover and exposed bedrock at some locations. Surface-wave testing indicated a mean $V_{S30}$ of $1{,}630 \pm 20$ m/s and bedrock $V_S$ values reaching approximately $3{,}150$ m/s. The corresponding 1D theoretical transfer functions remain near unity below approximately 10 Hz and predict amplification primarily at higher frequencies, indicating that the subsurface velocity structure alone cannot explain the observed low-frequency amplification. In contrast, two ambient-noise deployments showed amplification that was strongest near local ridge crests, decreased at slope and inland stations, occurred predominantly between 2 and 6 Hz, and was concentrated along directions approximately perpendicular to nearby ridge segments. At LD.PAL, repeated measurements collected four months apart showed similar amplification near 2 Hz, peak spectral ratios of approximately 1.6, and a consistent preferential direction. The earthquake-based S2S residuals also indicate amplification between approximately 0.6 and 2 Hz, overlapping the directional array response near 2 Hz. This low-frequency behavior is reproduced by neither the transfer functions nor the ergodic models. These observations identify topographic amplification, rather than one-dimensional subsurface response, as the primary mechanism contributing to the amplified motions at LD.PAL.

Taken together, the results demonstrate that two stations 38 km apart, with similar ground-motion residuals, required entirely different approaches to explain their behavior. An ergodic site term, which represents a site by the average response of many broadly similar sites, cannot distinguish between these mechanisms, and the parameters that explain one station carry almost no information about the other. At Caumsett, amplification is controlled primarily by the thickness and impedance structure of the deep sedimentary column, characteristics that are not adequately represented by $V_{S30}$ alone, but are well captured with sediment thickness and physiographic province. At Lamont-Doherty, considerable directional amplification occurs at a site that would otherwise be classified as hard rock with limited amplification. More broadly, the observed responses were associated with deep sediment structure and fundamental site frequency at Caumsett and with ridge topography and repeatable directional amplification at Lamont-Doherty. Targeted field characterization at strong-motion stations can identify these repeatable site effects, identify cases where non-1D effects (such as topography) may influence site response, and provide the observational basis needed to develop more physically informed, non-ergodic ground-motion models for the CEUS.

**Acknowledgments:** The authors thank Vincent Medina, Park Director of Caumsett State Historic Park Preserve, and James L. Davis of the Lamont-Doherty Earth Observatory of Columbia University, for facilitating access to the field sites.

**Funding:** This work was supported in part by faculty startup funds provided by the University of Rhode Island.

**Declaration of conflicting interests:** The authors declare no conflicts of interest.

**Supplemental material:** The electronic supplement provides additional detail on the surface-wave inversions at both sites. Section S1 presents the Caumsett State Historic Park results, comprising the complete 800-profile ensemble and its fit to the withheld Love-wave dispersion data (Figure S1), the median profile from each parameterization (Figure S2), the depth-dependent $V_S$ variability (Figure S3), the distribution of $V_{S30}$ (Figure S4), and the tabulated median profiles (Table S1). Section S2 presents the Lamont-Doherty Earth Observatory results, comprising the HVSR measurements (Figure S5), the construction of the experimental dispersion target (Figure S6), the inversion ensembles (Figure S7), the median $V_S$ profiles (Figure S8), and their tabulated values (Table S2). The field records and processed data from which these results were derived are available in the DesignSafe repository cited in the references.

## Supplementary Material for

# Explaining Ground-Motion Residuals at Two Strong-Motion Stations in Southeastern New York: Sediment Resonance and Topographic Amplification

Aser Abbas, Patrick Daniele, James Kaklamanos, Laurie Baise, Kyle Cannon, and Ellie Meyer

This supplement provides additional detail on the surface-wave testing results summarized in the main article. Section S1 presents the Caumsett State Historic Park results, including comparison with the withheld Love-wave dispersion data, median profiles from each parameterization, depth-dependent variability in the inverted shear-wave velocity ($V_S$) profiles, the distribution of the time-averaged shear-wave velocity in the upper 30 m ($V_{S30}$), and tabulated median profiles. Section S2 presents the Lamont-Doherty surface-wave testing results, including the horizontal-to-vertical spectral-ratio (HVSR) results, Rayleigh wave data dispersion trimming, ensemble and median inversion results, and tabulated median profiles.

## S1. Caumsett State Historic Park

The results presented in this section comprise the 100 lowest-misfit profiles from each of the eight best-performing parameterizations, for a total of 800 profiles, as described in Section 3.1 of the main article. Four parameterizations follow the layering-by-number approach (*ln* = 4, 6, 7, and 8) and four follow the layering-by-ratio approach (*lr* = 5.0, 6.0, 7.0, and 8.0). The misfit values reported in the figure legends are the standard-deviation-weighted misfits of Wathelet et al. (2004), computed from the combined Rayleigh-wave dispersion and Rayleigh-wave ellipticity-peak targets used in the inversion. Across the 800 selected profiles, these values range from 0.28 to 0.59.

This section includes the complete 800-profile ensemble and its fit to the withheld Love-wave dispersion data (Figure S1), the median profile from each parameterization (Figure S2), the depth-dependent $V_S$ variability (Figure S3), the distribution of $V_{S30}$ (Figure S4), and the tabulated median profiles (Table S1).

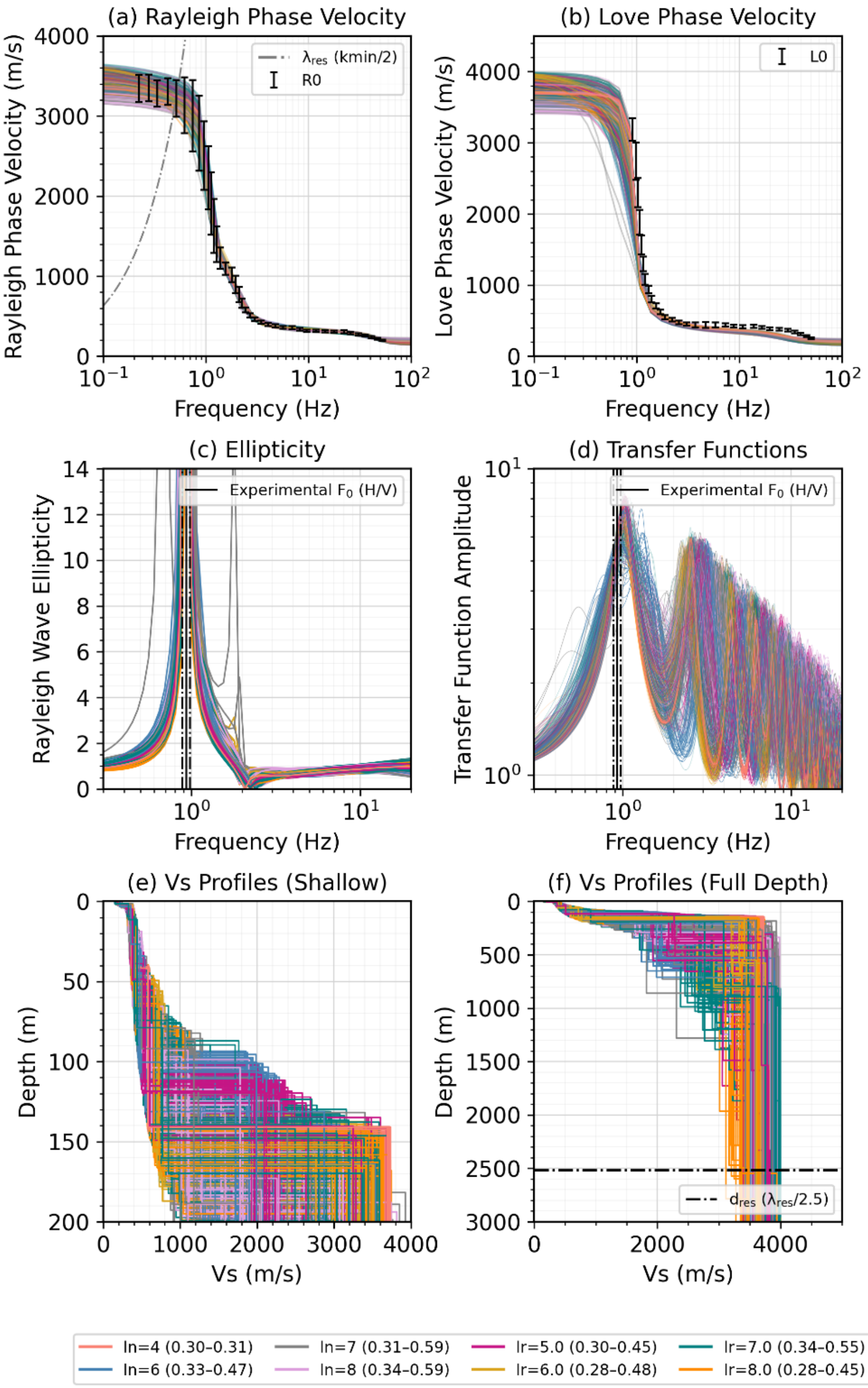

**Figure S1.** Complete inversion results at Caumsett State Historic Park for the 800 selected shear-wave velocity profiles, comprising the 100 lowest-misfit profiles from each of the eight parameterizations. (a) Experimental fundamental-mode Rayleigh-wave dispersion data (R0, shown as mean and plus or minus one standard deviation) and the theoretical Rayleigh-wave dispersion curves computed from the selected profiles. The gray dash-dotted curve indicates the theoretical array-resolution limit. (b) Experimental fundamental-mode Love-wave dispersion data (L0) and the theoretical Love-wave dispersion curves computed from the same profiles. The Love-wave data were withheld from the inversion and are used here as an independent check on the inverted velocity profiles. (c) Theoretical fundamental-mode Rayleigh-wave ellipticity curves and (d) one-dimensional linear theoretical transfer functions computed from the selected profiles. In (c) and (d), the solid vertical line indicates the measured spatially weighted lognormal median peak frequency, $LM_{f0}$ = 0.93 Hz, and the dash-dotted vertical lines indicate its plus or minus one natural-log standard deviation bounds ($\sigma_{ln,f0}$ = 0.05). (e) Selected profiles over the upper 200 m and (f) to a depth of 3,000 m; the horizontal dash-dotted line in (f) indicates the resolution depth, $d_{res} = \lambda_{res}/2.5 \approx$ 2,500 m. Colors identify the parameterization (*ln*, layering by number; *lr*, layering by ratio), and the values in parentheses give the range of misfit values among the 100 profiles selected for each parameterization.

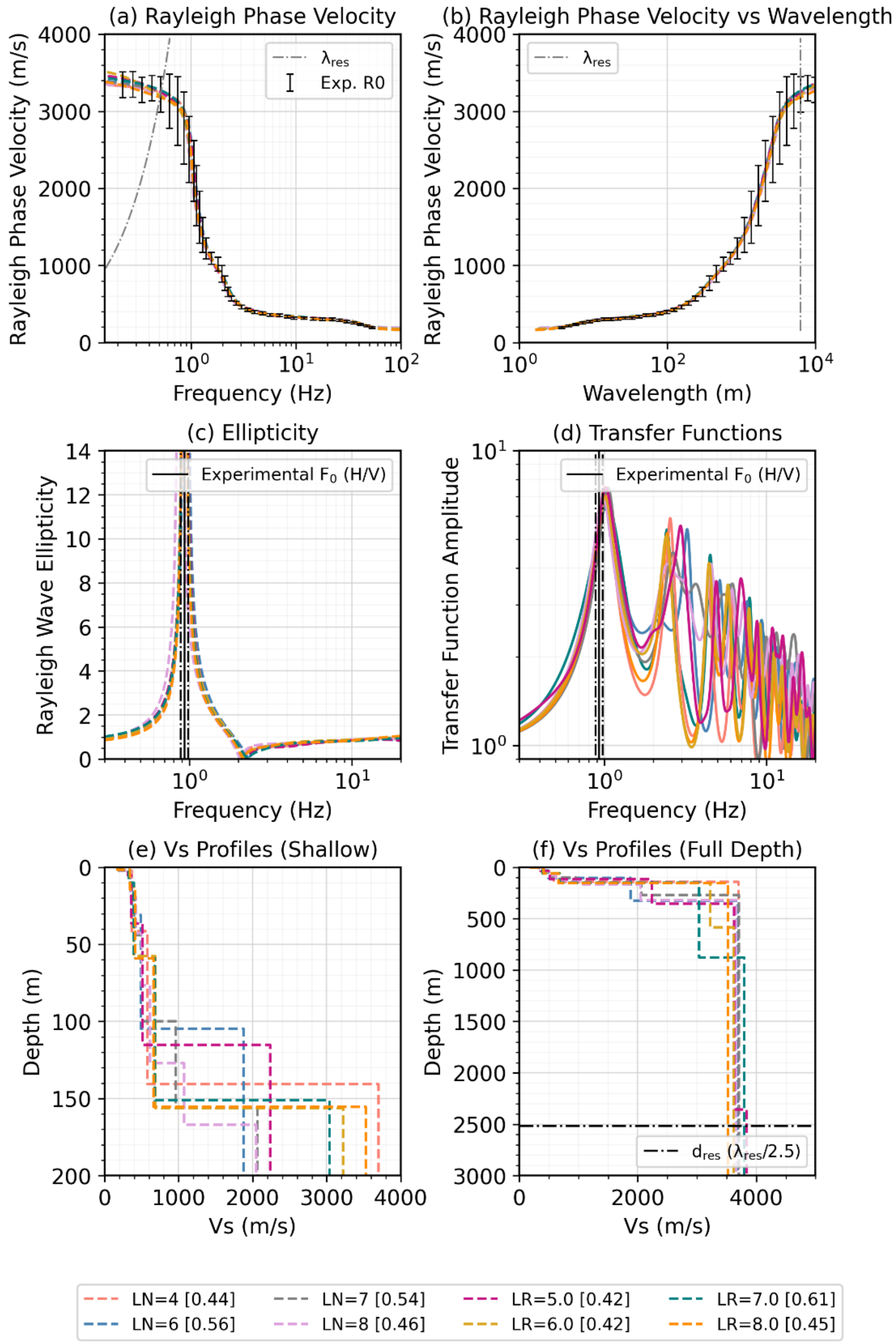

**Figure S2.** Inversion results at Caumsett State Historic Park for the eight median shear-wave velocity profiles, one from each parameterization. (a) Experimental fundamental-mode Rayleigh-wave dispersion data and the theoretical dispersion curves computed from the median profiles, plotted against frequency, and (b) the same curves plotted against wavelength. The gray dash-dotted line in (a) and (b) indicates the theoretical array-resolution limit. (c) Theoretical fundamental-mode Rayleigh-wave ellipticity curves and (d) one-dimensional linear theoretical transfer functions computed from the median profiles, with $LM_{f0}$ and its bounds shown as in Figure S1. (e) Median profiles over the upper 200 m and (f) to a depth of 3,000 m; the horizontal dash-dotted line in (f) indicates the resolution depth, as in Figure S1. The value in brackets in the legend is the misfit of each median profile. Because the median profile is derived from the suite of 100 profiles rather than drawn from it, its misfit is not constrained to fall within the misfit range reported for that parameterization in Figure S1.

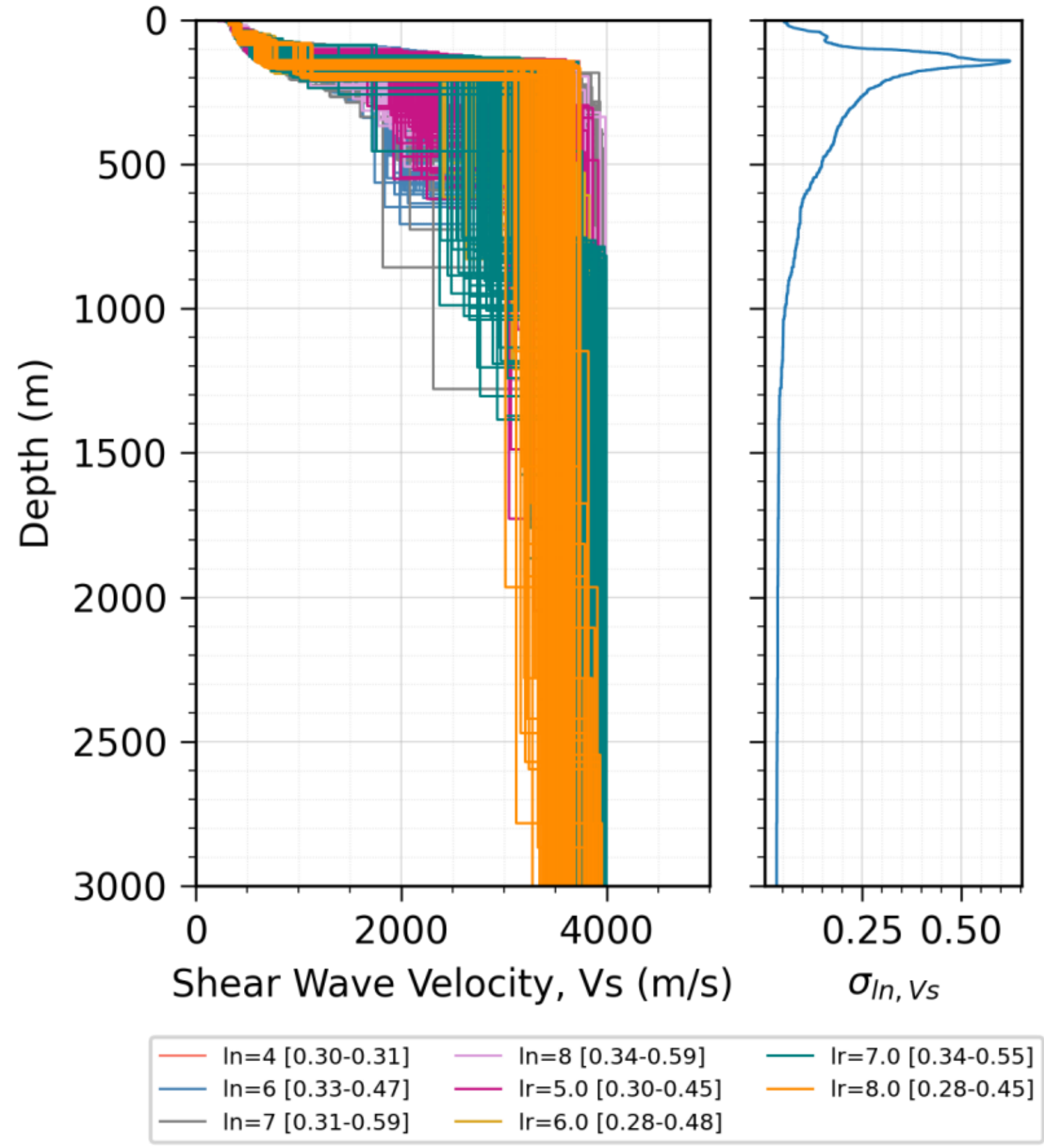

**Figure S3.** Variability of the inverted shear-wave velocity with depth at Caumsett State Historic Park. Left: the 800 selected profiles, colored by parameterization. Right: the natural-log standard deviation of shear-wave velocity, $\sigma_{ln,Vs}$, computed across all 800 profiles at each depth. $\sigma_{ln,Vs}$ is approximately 0.05 to 0.09 within the upper 30 m, increases to a maximum of approximately 0.6 near 145 m depth, and decreases to approximately 0.04 below about 1,000 m. The maximum is associated with the range of depths at which the individual profiles place the impedance contrast rather than with variability in the velocity of the sediments or the bedrock.

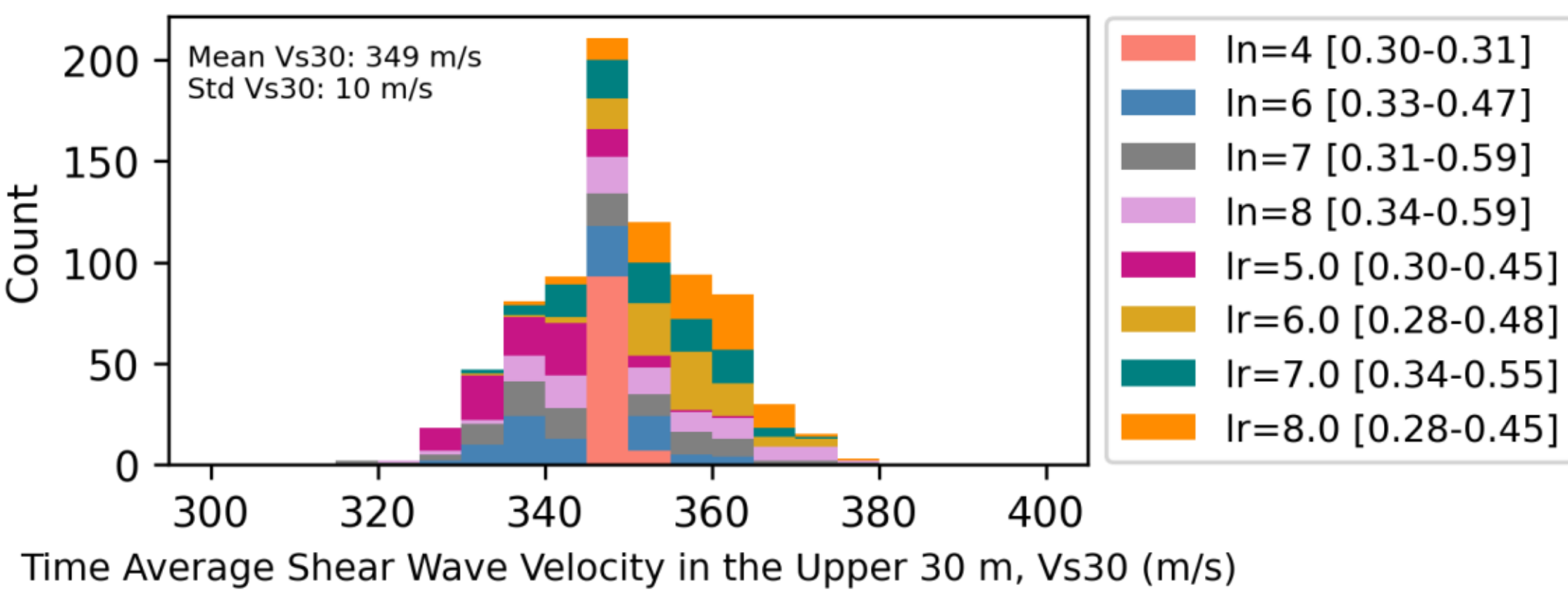

**Figure S4.** Distribution of the time-averaged shear-wave velocity in the upper 30 m, $V_{S30}$, for the 800 selected profiles at Caumsett State Historic Park. Values are binned in 5 m/s intervals and stacked by parameterization. The bracketed values in the legend give the range of Rayleigh-wave dispersion misfits among the 100 selected profiles for each parameterization. The mean $V_{S30}$ is 349 m/s, with a standard deviation of 10 m/s.

**Table S1.** Median shear-wave velocity profiles at Caumsett State Historic Park for the four layering-by-number and the four layering-by-ratio parameterizations. Each profile is given as depth and shear-wave velocity pairs. Rows at 2,513 m were inserted to identify the resolution limit. Shaded values lie below the resolution depth, $d_{res}$ = 2,513 m, where the profiles are constrained by less reliable dispersion data and should be used with caution. Velocities below $d_{res}$ are reported because, although less certain, they provide better guidance on the deep velocity structure than an assumed value.

| **Layering by number, ln = 4** | | **Layering by number, ln = 6** | | **Layering by number, ln = 7** | | **Layering by number, ln = 8** | |
|---|---|---|---|---|---|---|---|
| **Depth (m)** | **$V_S$ (m/s)** | **Depth (m)** | **$V_S$ (m/s)** | **Depth (m)** | **$V_S$ (m/s)** | **Depth (m)** | **$V_S$ (m/s)** |
| 0.0 | 205 | 0.0 | 202 | 0.0 | 191 | 0.0 | 201 |
| 2.0 | 205 | 2.2 | 202 | 2.0 | 191 | 2.2 | 201 |
| 2.0 | 367 | 2.2 | 347 | 2.0 | 351 | 2.2 | 353 |
| 41.5 | 367 | 8.7 | 347 | 17.9 | 351 | 19.3 | 353 |
| 41.5 | 580 | 8.7 | 367 | 17.9 | 409 | 19.3 | 423 |
| 140.7 | 580 | 30.9 | 367 | 43.9 | 409 | 42.6 | 423 |
| 140.7 | 3699 | 30.9 | 489 | 43.9 | 516 | 42.6 | 502 |
| 2513 | 3699 | 104.7 | 489 | 99.8 | 516 | 76.7 | 502 |
| 2513 | 3699 | 104.7 | 1877 | 99.8 | 965 | 76.7 | 614 |
| 4000.0 | 3699 | 323.7 | 1877 | 156.3 | 965 | 127.1 | 614 |
| - | - | 323.7 | 3683 | 156.3 | 2064 | 127.1 | 1077 |
| - | - | 2513 | 3683 | 268.2 | 2064 | 167.2 | 1077 |
| - | - | 2513 | 3683 | 268.2 | 3710 | 167.2 | 2048 |
| - | - | 4000.0 | 3683 | 2513 | 3710 | 321.1 | 2048 |
| - | - | - | - | 2513 | 3710 | 321.1 | 3659 |
| - | - | - | - | 4000.0 | 3710 | 2513 | 3659 |
| - | - | - | - | - | - | 2513 | 3659 |
| - | - | - | - | - | - | 4000.0 | 3659 |
| | | | | | | | |
| **Layering by ratio, lr = 5.0** | | **Layering by ratio, lr = 6.0** | | **Layering by ratio, lr = 7.0** | | **Layering by ratio, lr = 8.0** | |
| **Depth (m)** | **$V_S$ (m/s)** | **Depth (m)** | **$V_S$ (m/s)** | **Depth (m)** | **$V_S$ (m/s)** | **Depth (m)** | **$V_S$ (m/s)** |
| 0.0 | 183 | 0.0 | 186 | 0.0 | 173 | 0.0 | 175 |
| 1.7 | 183 | 1.8 | 186 | 1.6 | 173 | 1.5 | 175 |
| 1.7 | 313 | 1.8 | 339 | 1.6 | 326 | 1.5 | 345 |
| 5.1 | 313 | 10.3 | 339 | 9.9 | 326 | 14.2 | 345 |
| 5.1 | 362 | 10.3 | 394 | 9.9 | 396 | 14.2 | 412 |
| 36.3 | 362 | 57.6 | 394 | 59.0 | 396 | 58.9 | 412 |
| 36.3 | 515 | 57.6 | 683 | 59.0 | 677 | 58.9 | 663 |
| 115.2 | 515 | 156.2 | 683 | 151.1 | 677 | 155.4 | 663 |
| 115.2 | 2238 | 156.2 | 3223 | 151.1 | 3037 | 155.4 | 3526 |
| 354.2 | 2238 | 585.7 | 3223 | 877.3 | 3037 | 2513 | 3526 |
| 354.2 | 3630 | 585.7 | 3619 | 877.3 | 3797 | 4000.0 | 3526 |
| 2357.9 | 3630 | 2513 | 3619 | 2513 | 3797 | - | - |
| 2357.9 | 3839 | 2513 | 3619 | 2513 | 3797 | - | - |
| 2513 | 3839 | 3896.2 | 3619 | 4000.0 | 3797 | - | - |
| 2513 | 3839 | 3896.2 | 3942 | - | - | - | - |
| 4000.0 | 3839 | 4000.0 | 3942 | - | - | - | - |

## S2. Lamont-Doherty Earth Observatory

The subsurface $V_S$ at Lamont-Doherty Earth Observatory was evaluated using multichannel analysis of surface waves (MASW) and circular microtremor array measurements (MAM) with diameters of approximately 50 m (C50) and 300 m (C300). Utilities and research infrastructure affected the original MASW and C50 locations, so these measurements were relocated. Even at the relocated position, the C50 array did not yield consistent Rayleigh-wave dispersion data; the observations remained scattered over a wide velocity range at each frequency and did not define a coherent fundamental mode (Figure S6a). This behavior is attributed to buried utilities remaining within the array footprint, and the C50 data were therefore excluded from the inversion target. The MASW result provides a well-resolved profile at the location of the acquisition line, whereas the C300 array spans a large portion of the site and averages local changes in the very thin surficial sediment veneer. The experimental target combines the short-wavelength MASW observations with the longer-wavelength C300 observations, and the C300 dispersion data provide the primary site-scale constraint on the shear-wave velocity structure.

The HVSR curves vary among the C50 and C300 arrays and the two additional Lamont deployments (LamTA01 and LamTA02; Figure S5). The identified median peaks occur at high frequencies and vary across the site. Some curves, predominantly from the C50 array, also show elevated HVSR amplitudes below approximately 1.5 Hz. The C50 array was located over buried utilities and adjacent research infrastructure, the same conditions that prevented a consistent C50 dispersion estimate, so these records are the least representative of site-scale conditions. These amplitudes were not identified as peaks, as they are spatially inconsistent; comparable amplitudes are not observed in the LamTA01 and LamTA02 deployments, which sampled a much larger portion of the site. The peaks are therefore not interpreted as a site-wide resonance. Taken together, the high and spatially variable peak frequencies are consistent with a sediment veneer that is both thin and laterally variable, rather than a laterally continuous layer producing a single site-wide fundamental resonance. The surface-wave results indicate that most of the velocity variation is restricted to approximately the upper 2 m, below which the profiles converge toward a nearly constant high bedrock velocity. Consequently, a site-wide fundamental frequency was not targeted during the inversion, and the HVSR curves were not treated as Rayleigh-wave ellipticity constraints.

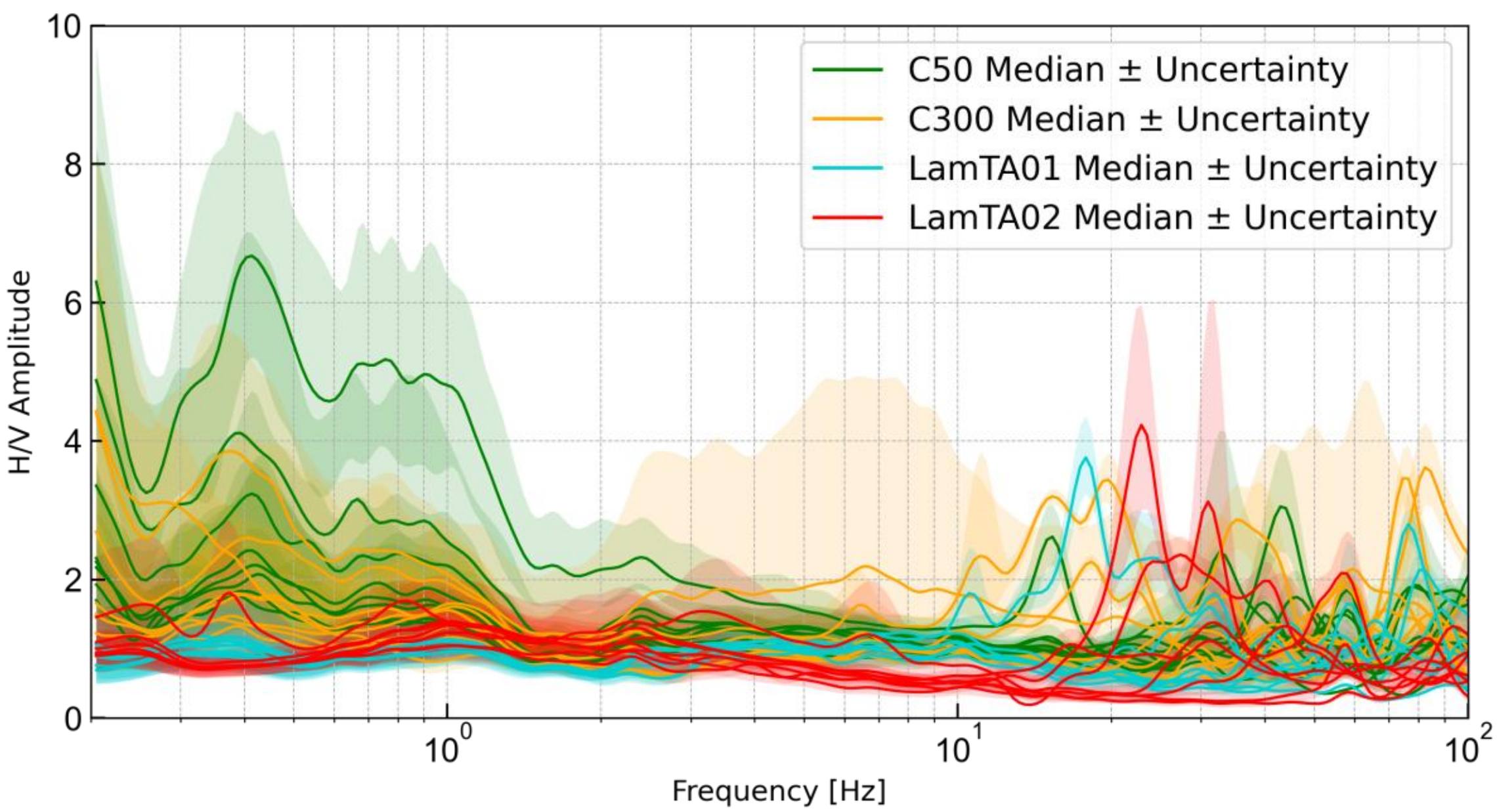

**Figure S5.** Horizontal-to-vertical spectral ratios (HVSR) at Lamont-Doherty Earth Observatory. Individual median curves are shown for each station from the C50 and C300 arrays and the LamTA01 and LamTA02 deployments. The shaded bands span one lognormal standard deviation below and above the corresponding median curve.

Figure S6 documents the construction of the experimental Rayleigh wave phase velocity dispersion target. The untrimmed observations include the MASW, C50, and C300 results together with the corresponding array-resolution limits. The retained target combines the MASW observations at shorter wavelengths with the C300 observations that sample a larger portion of the site at longer wavelengths.

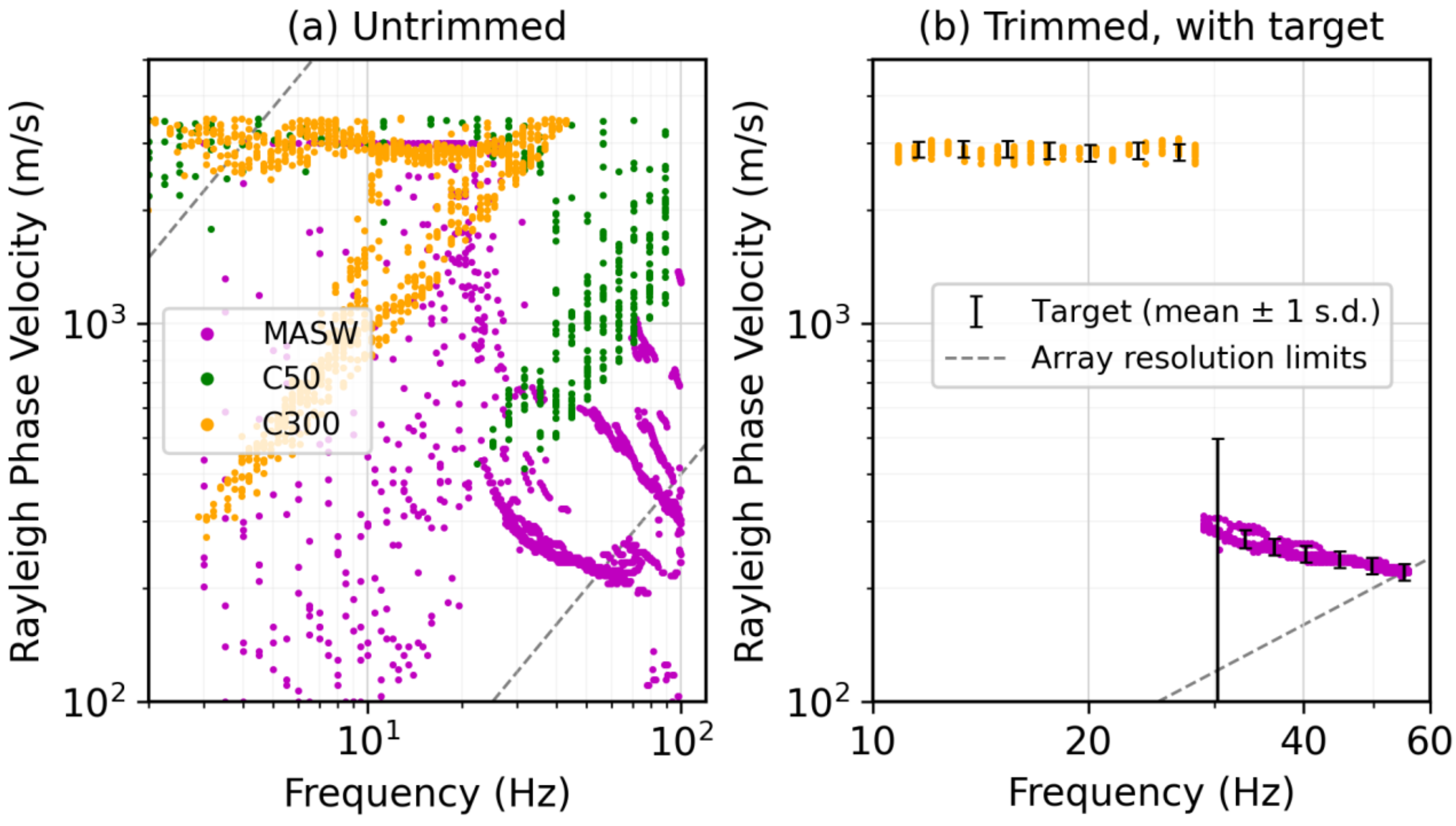

**Figure S6.** Rayleigh-wave phase-velocity dispersion-data selection for the Lamont-Doherty inversion. (a) Untrimmed MASW, C50, and C300 Rayleigh-wave phase-velocity observations. Dashed lines indicate the array-resolution limits. (b) Experimental fundamental-mode Rayleigh-wave target retained for inversion, shown as the mean plus or minus one standard deviation. The retained target combines the short-wavelength MASW observations with the longer-wavelength C300 observations.

The inversion used eight parameterizations that varied the number of layers (ln = 2, 3, 5, and 7) and the layer-thickness ratio (lr = 1.5, 3.0, 5.0, and 7.0). Figure S7 shows the 100 lowest-misfit models from each parameterization and their predicted fundamental-mode Rayleigh-wave dispersion curves. All parameterizations reproduce the experimental target over the retained frequency and wavelength ranges. Model variability is greatest within the upper approximately 2 m, where the thin sediment veneer is resolved, and decreases rapidly beneath it as the profiles converge toward the high-velocity diabase. The depth-resolution limit is approximately 300 m.

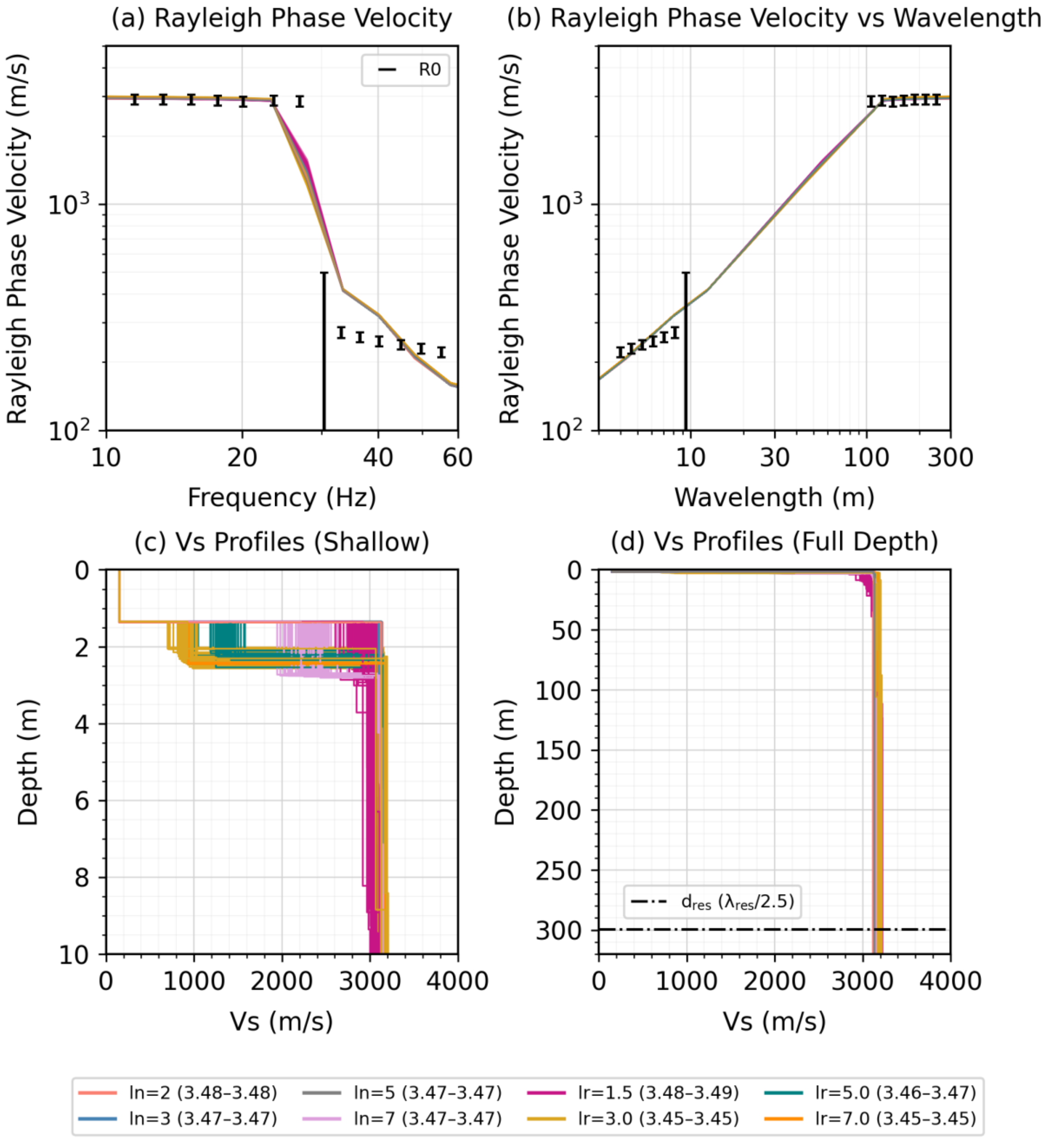

**Figure S7.** Lamont-Doherty inversion ensembles comprising the 100 lowest-misfit models from each of the eight parameterizations. (a) Theoretical fundamental-mode Rayleigh-wave dispersion curves as a function of frequency with the experimental $R_0$ target. (b) The same curves as a function of wavelength. (c) Shear-wave velocity profiles over the upper 10 m. (d) Profiles to 320 m depth. The horizontal dashed line in panel (d) marks the approximate depth-resolution limit, $d_{res}$. Bracketed values in the shared legend give the misfit range for each parameterization.

Figure S8 compares the median models from the eight parameterizations. All show a rapid transition within the upper 2 m to a high and comparatively uniform bedrock shear-wave velocity. Agreement at depth reflects the longer-wavelength C300 constraint. Shallow differences reflect parameterization uncertainty and the local MASW constraint, while the spatially variable HVSR measurements independently support lateral variability within the veneer. Table S2 provides the numerical profiles.

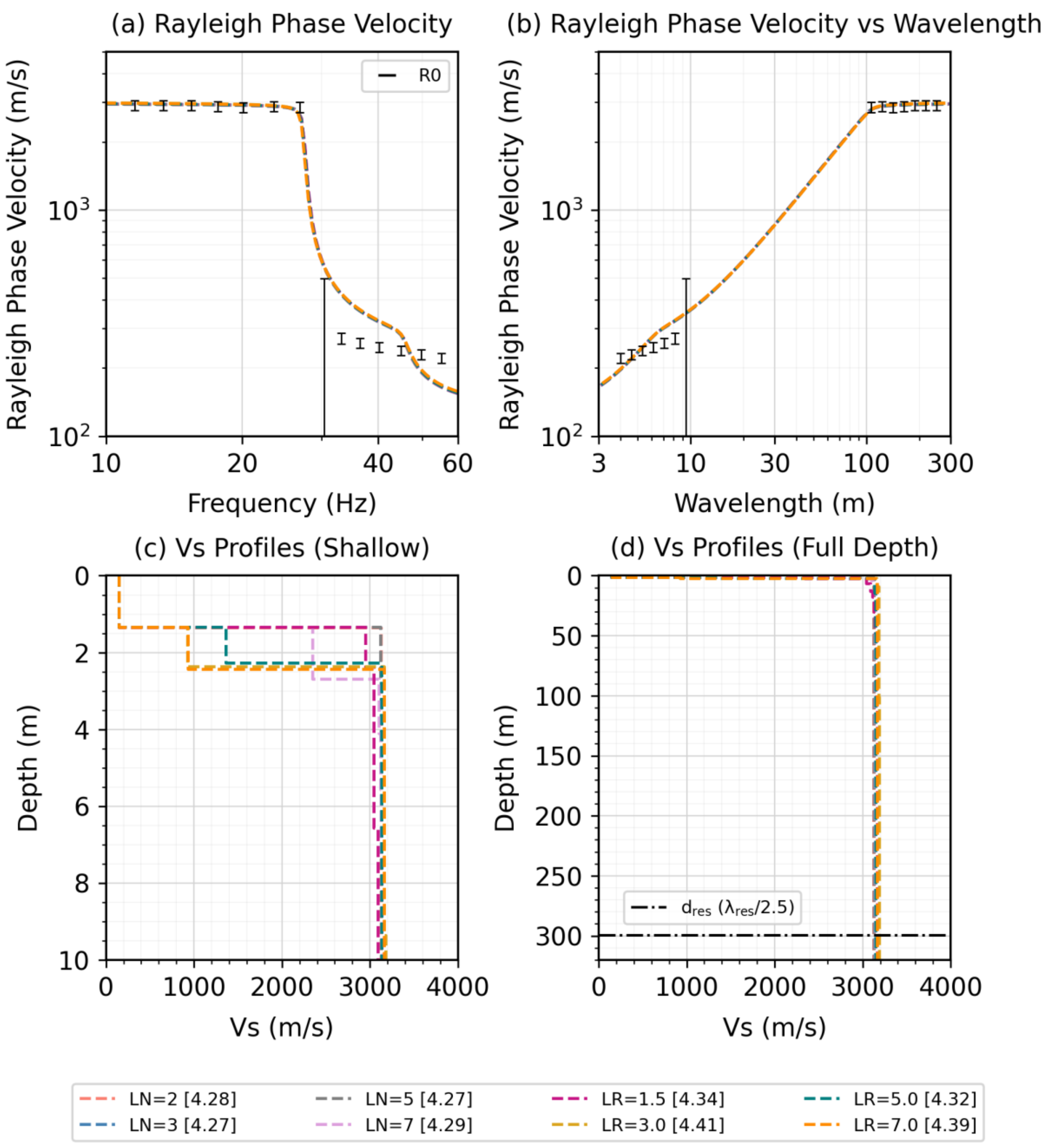

**Figure S8.** Median Lamont-Doherty inversion results for the eight parameterizations. (a) Fundamental-mode Rayleigh-wave dispersion curves as a function of frequency with the experimental R0 target. (b) The same curves as a function of wavelength. (c) Median shear-wave velocity profiles over the upper 10 m. (d) Median profiles to 320 m depth. The horizontal dashed line in panel (d) marks the approximate depth-resolution limit, $d_{res}$. Bracketed values in the shared legend give the misfit of the median model for each parameterization. Because each median profile is derived from the suite of 100 models rather than selected from it, its misfit is not constrained to fall within the range shown in Figure S7.

**Table S2.** Median shear-wave velocity profiles at Lamont-Doherty Earth Observatory for the four layering-by-number and the four layering-by-ratio parameterizations. Each profile is given as depth and $V_S$ pairs. The shaded rows at 300 m were inserted to identify the approximate resolution depth.

| Layering by number, ln = 2 | | Layering by number, ln = 3 | | Layering by number, ln = 5 | | Layering by number, ln = 7 | |
|---|---|---|---|---|---|---|---|
| **Depth (m)** | **$V_S$ (m/s)** | **Depth (m)** | **$V_S$ (m/s)** | **Depth (m)** | **$V_S$ (m/s)** | **Depth (m)** | **$V_S$ (m/s)** |
| 0.0 | 150 | 0.0 | 149 | 0.0 | 149 | 0.0 | 150 |
| 1.4 | 150 | 1.3 | 149 | 1.3 | 149 | 1.3 | 150 |
| 1.4 | 3126 | 1.3 | 3120 | 1.3 | 3123 | 1.3 | 2349 |
| 300.0 | 3126 | 2.7 | 3120 | 2.7 | 3123 | 2.7 | 2349 |
| 300.0 | 3126 | 2.7 | 3132 | 2.7 | 3126 | 2.7 | 3104 |
| - | - | 300.0 | 3132 | 4.1 | 3126 | 4.1 | 3104 |
| - | - | 300.0 | 3132 | 4.1 | 3129 | 4.1 | 3123 |
| - | - | - | - | 5.6 | 3129 | 5.6 | 3123 |
| - | - | - | - | 5.6 | 3132 | 5.6 | 3129 |
| - | - | - | - | 300.0 | 3132 | 7.4 | 3129 |
| - | - | - | - | 300.0 | 3132 | 7.4 | 3138 |
| - | - | - | - | - | - | 9.8 | 3138 |
| - | - | - | - | - | - | 9.8 | 3145 |
| - | - | - | - | - | - | 300.0 | 3145 |
| - | - | - | - | - | - | 300.0 | 3145 |
| | | | | | | | |
| **Layering by ratio, lr = 1.5** | | **Layering by ratio, lr = 3.0** | | **Layering by ratio, lr = 5.0** | | **Layering by ratio, lr = 7.0** | |
| **Depth (m)** | **$V_S$ (m/s)** | **Depth (m)** | **$V_S$ (m/s)** | **Depth (m)** | **$V_S$ (m/s)** | **Depth (m)** | **$V_S$ (m/s)** |
| 0.0 | 150 | 0.0 | 152 | 0.0 | 150 | 0.0 | 152 |
| 1.3 | 150 | 1.3 | 152 | 1.3 | 150 | 1.3 | 152 |
| 1.3 | 2953 | 1.3 | 928 | 1.3 | 1366 | 1.3 | 933 |
| 2.4 | 2953 | 2.4 | 928 | 2.3 | 1366 | 2.4 | 933 |
| 2.4 | 3046 | 2.4 | 3162 | 2.3 | 3135 | 2.4 | 3167 |
| 6.6 | 3046 | 9.4 | 3162 | 14.4 | 3135 | 18.1 | 3167 |
| 6.6 | 3090 | 9.4 | 3183 | 14.4 | 3148 | 18.1 | 3170 |
| 13.0 | 3090 | 101.5 | 3183 | 89.5 | 3148 | 300.0 | 3170 |
| 13.0 | 3116 | 101.5 | 3191 | 89.5 | 3152 | 300.0 | 3170 |
| 18.0 | 3116 | 300.0 | 3191 | 300.0 | 3152 | - | - |
| 18.0 | 3132 | 300.0 | 3191 | 300.0 | 3152 | - | - |
| 30.3 | 3132 | - | - | - | - | - | - |
| 30.3 | 3145 | - | - | - | - | - | - |
| 47.7 | 3145 | - | - | - | - | - | - |
| 47.7 | 3154 | - | - | - | - | - | - |
| 109.4 | 3154 | - | - | - | - | - | - |
| 109.4 | 3170 | - | - | - | - | - | - |
| 300.0 | 3170 | - | - | - | - | - | - |
| 300.0 | 3170 | - | - | - | - | - | - |